# Imaging Chern mosaic and Berry-curvature magnetism in magic-angle graphene

S. Grover[1†], M. Bocarsly[1†], A. Uri[1†], P. Stepanov[2], G. Di Battista[2], I. Roy[1], J. Xiao[1], A. Y. Meltzer[1], Y. Myasoedov[1], K. Pareek[1], K. Watanabe[3], T. Taniguchi[4], B. Yan[1], A. Stern[1], E. Berg[1], D. K. Efetov[2] and E. Zeldov[1]*

Charge carriers in magic angle graphene come in eight flavors described by a combination of their spin, valley, and sublattice polarizations [1–4]. When the inversion and time reversal symmetries are broken by the substrate or by strong interactions, the degeneracy of the flavors can be lifted and their corresponding bands can be filled sequentially [5–7]. Due to their non-trivial band topology and Berry curvature, each of the bands is classified by a topological Chern number [8–12], leading to the quantum anomalous Hall [13–15] and Chern insulator states [7,8,16–19] at integer fillings $v$ of the bands. It has been recently predicted, however, that depending on the local atomic-scale arrangements of the graphene and the encapsulating hBN lattices, rather than being a global topological invariant, the Chern number $C$ may become position dependent, altering transport and magnetic properties of the itinerant electrons [20–23]. Using scanning superconducting quantum interference device on a tip (SQUID-on-tip) [24], we directly image the nanoscale Berry-curvature-induced equilibrium orbital magnetism, the polarity of which is governed by the local Chern number, and detect its two constituent components associated with the drift and the self-rotation of the electronic wave packets [25]. At $v = 1$, we observe local zero-field valley-polarized Chern insulators forming a mosaic of microscopic patches of $C = -1$, 0, or 1. Upon further filling, we find a first-order phase transition due to recondensation of electrons from valley $K$ to $K'$, which leads to irreversible flips of the local Chern number and the magnetization, and to the formation of valley domain walls giving rise to hysteretic global anomalous Hall resistance. The findings shed new light on the structure and dynamics of topological phases and call for exploration of the controllable formation of flavor domain walls and their utilization in twistronic devices.

___________________________

[1]Department of Condensed Matter Physics, Weizmann Institute of Science, Rehovot 7610001, Israel

[2]ICFO - Institut de Ciencies Fotoniques, The Barcelona Institute of Science and Technology, Castelldefels, Barcelona, 08860, Spain

[3]Research Center for Functional Materials, National Institute for Materials Science, 1-1 Namiki, Tsukuba 305-0044, Japan

[4]International Center for Materials Nanoarchitectonics, National Institute for Materials Science, 1-1 Namiki, Tsukuba 305-0044, Japan

[†]These authors contributed equally to this work

*eli.zeldov@weizmann.ac.il

Magic angle twisted bilayer graphene (MATBG) exhibits a unique combination of strong electron correlations with band topology that is not readily accessible in other materials, giving rise to a wealth of emergent phenomena in a single system. Bringing the full potential of these traits to light entails two elements. The first is gapping of the Dirac cones that can be attained by application of magnetic fields or by structural modifications, such as alignment with hBN. With a gapped dispersion, the Berry curvature, rather than being singular at the Dirac point, becomes extended over the band, giving rise to energy dependent orbital magnetism [25–27]; and the band topology, rather than being a subtle property, is expressed profoundly in terms of nontrivial Chern gaps [8–12,28]. The second is lifting the time reversal symmetry, otherwise bands related by time reversal act to cancel the effects of one another. The unique combination in MATBG of gapped Dirac cones and spontaneous degeneracy lifting due to strong interactions leads to Chern insulator phases [7,13–19,29–31] and to orbital ferromagnetic states characterized by the quantum anomalous Hall (QAH) effect at full filling of a Chern band [13–15,30,32].

The Chern insulators in MATBG have been revealed by transport [7,16–19,31], STM [33,34], and compressibility measurements [29]. These experiments provide an unambiguous evidence of gapped degeneracy-lifted bands at integer fillings $\nu$ that are characterized by a nontrivial Chern number $C \in \mathbb{Z}$, which defines their topological state. Recent theoretical works [20–23], however, have proposed a new perspective on this fundamental concept: rather than being a fixed global property, a spatially varying $C$ may occur due to position dependent substrate potential. This additional degree of freedom opens a new platform for construction and manipulation of previously unconsidered confined topological states and gapless modes. In this work we provide the first experimental test of this hypothesis.

The Berry-curvature-induced orbital magnetism, even though studied extensively theoretically [25,27,35–37] and being the main source of the anomalous Hall resistivity, has evaded direct experimental observation due to lack of sufficiently sensitive probes. Moreover, the anomalous Hall effect (AHE), indicative of magnetism, has been reported in just a few MATBG samples [13–15,19]. Of these samples, only one was not intentionally aligned to hBN [19], and showed both AHE at $\nu = 1$ and superconductivity. All hBN aligned samples showed AHE or QAH at $\nu = 3$, with no evidence of superconductivity. Recently, the first direct measurement of the local magnetization hysteresis in the gapped QAH state in MATBG aligned to hBN has been reported [15]. Upon sweeping the magnetic field at fixed $\nu = 3$, it showed a hysteretic magnetization reversal of about 3 $\mu_B$ per charge carrier, providing strong evidence for the orbital nature of the magnetization. Here we image the continuous evolution of the equilibrium Berry-curvature-induced orbital magnetism as a function of filling factor, through both the metallic and the gapped states, in the nonaligned MATBG sample of Ref. [19]. The revealed dynamics shows a complicated magnetization map that changes locally in magnitude and sign as a function of carrier density.

**Transport measurements**

The hBN encapsulated MATBG sample with a twist angle $\theta \cong 1.08°$ was fabricated using the cut-and-stack technique (Methods). Transport measurements of $R_{xx}$ and $R_{yx}$ (Figs. 1a,b) were performed in a Hall bar geometry (Fig. 2b) at a temperature $T = 300$ mK as a function of applied perpendicular magnetic field $B_a$ and filling factor $\nu = 4n/n_s = 4C_{bg}V_{bg}^{dc}/n_s$, where $V_{bg}^{dc}$ is the dc voltage applied to the graphite backgate, $C_{bg}$ is the backgate capacitance, $n$ is the carrier density, and $n_s$ corresponds to a full flat band (four electrons per moiré cell). Characteristic transport features of high quality MATBG are evident including superconductivity at $\nu = -2 - \delta$, correlated and high-field Chern insulating states, and Landau fans (Extended Data Fig. 2). Notably, near $\nu = 1$ and at low $B_a$, a non-trivial $R_{yx}$ behavior is observed (Figs. 1c-g), which indicates the presence of orbital magnetism and anomalous Hall effect (AHE) without full quantization, similar to previous reports [13,19]. In particular, we resolve a pronounced hysteresis in $R_{yx}$ which displays Barkhausen jumps upon sweeping $B_a$ (Figs. 1e,f), as well as $\nu$ (Figs. 1c,d). In addition, we have performed minor hysteresis loops by sweeping $\nu$ back and forth from $\nu = 0.55$ to $\nu_{max}$ and gradually increasing $\nu_{max}$, as shown in Fig. 1g for $B_a = 46$ mT (see Supplementary Video 1). For $\nu_{max} \leq 1$, $R_{yx}$ is reversible, while at higher filling hysteresis sets in and grows rapidly. Notably, the full size of the hysteresis loop is attained only when $\nu_{max}$ exceeds 1.25, which is significantly larger than $\nu = 1$.

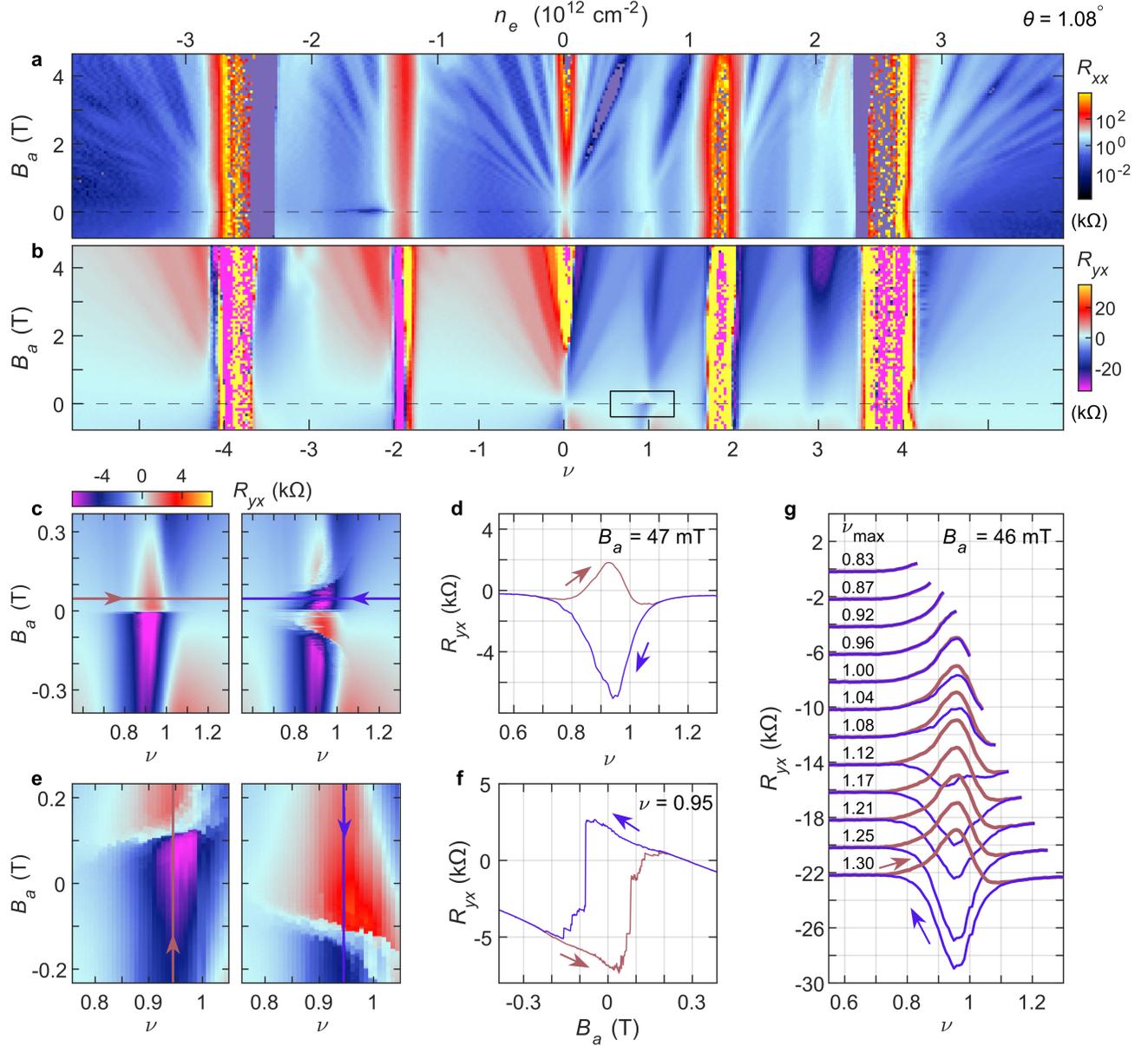

**Fig. 1. Transport characteristics of MATBG. a-b**, Measured $R_{xx}$ (**a**) and $R_{yx}$ (**b**) at $T = 300$ mK vs. carrier concentration and the applied out-of-plane magnetic field $B_a$ displaying superconductivity, correlated insulating states and Landau fans. **c**, $R_{yx}$ in the vicinity of $\nu = 1$ at low $B_a$ (black rectangle in (**b**)) displaying pronounced hysteresis upon sweeping $\nu$ up (left panel) and down (right). **d**, Line cuts along the red and blue arrows in **c** at $B_a = 47$ mT. **e**, $R_{yx}$ displaying hysteresis upon sweeping $B_a$ up (left panel) and down (right). **f**, Line cuts along the red and blue arrows in (**e**) at $\nu = 0.95$. **g**, Minor hysteresis loops of $R_{yx}$ upon sweeping $\nu$ from 0.6 to $\nu_{max}$ (red) and back (blue) at $B_a = 46$ mT while incrementing $\nu_{max}$ from 0.83 to 1.30 as indicated (see Supplementary Video 1). The consecutive curves are offset by $-2$ kΩ for clarity.

**Magnetic imaging**

To study the local magnetism we utilize a scanning superconducting quantum interference device fabricated on the apex of a sharp pipette (SQUID-on-tip, SOT) [24]. An Indium SOT [38] of diameter $d \cong 180$ nm and a field sensitivity of down to 10 nT/Hz$^{1/2}$, is scanned at a height of $h \cong 160$ nm above the sample surface at $T = 300$ mK in presence of small applied $B_a \cong 50$ mT (Methods). By applying $V_{bg}^{dc}$ to the graphite backgate (Fig. 2a) we set the filling factor near $\nu = 1$ and add a small $ac$ $V_{bg}^{ac} = 20$ mV (peak-to-peak, square wave) with a frequency of $f = 5$ to 6 kHz, which modulates the carrier density by $n^{ac}$ corresponding to filling factor modulation of $\nu^{ac} = 0.083$, and image the resulting $B_z^{ac}(x,y) = n^{ac}(dB_z/dn) = \nu^{ac}(dB_z/d\nu)$ across the sample. This signal reflects the change in the local magnetization due to a small change in the carrier density in the MATBG. We begin with line scans along the black dotted line in Fig. 2b. In Fig. 2f the system is initialized at $\nu = 0$ and then $B_z^{ac}(x, \nu_\uparrow)$ is measured in the vicinity of $\nu = 1$ as $\nu$ is incremented up, $\nu_\uparrow$. A pronounced $B_z^{ac}$ is observed around $\nu = 1$ which commences well below $\nu = 1$ and extends significantly above it. Figure 2g shows the $B_z^{ac}(x, \nu_\downarrow)$ which is attained upon initialization at $\nu = 2$ and decrementing the filling factor down, $\nu_\downarrow$. By subtracting Figs. 2f and 2g, we plot in Fig. 2h the $\Delta B_z^{ac}(x,\nu) = (B_z^{ac}(x,\nu_\uparrow) - B_z^{ac}(x,\nu_\downarrow))/2$, which reveals that away from $\nu = 1$ the $B_z^{ac}$ is reversible, while in the vicinity of $\nu = 1$ a significant fraction of the locations along the scan line show magnetization hysteresis.

Next, we carry out area scans measuring $B_z^{ac}(x,y)$ in the orange rectangle in Fig. 2b, following the same $\nu_\uparrow$ and $\nu_\downarrow$ procedure and focus closer on the vicinity of $\nu = 1$. The resulting $B_z^{ac}(x,y,\nu_\uparrow)$ and $B_z^{ac}(x,y,\nu_\downarrow)$ at $\nu = 0.966$ in Figs. 2c,d reveal a complex pattern of positive and negative $B_z^{ac}$. Figure 2e shows $\Delta B_z^{ac}$ revealing the presence of magnetization hysteresis in substantial parts of the sample, whereas other areas are completely reversible at this $\nu$. Supplementary Video 2 presents an intricate evolution of $B_z^{ac}$ upon sweeping $\nu_\uparrow$ and $\nu_\downarrow$ in the range of $\nu = 0.737$ to $1.174$.

By preforming a numerical inversion of $B_z^{ac}(x,y)$, we reconstruct a map of local differential magnetization $m_z(x,y) = dM_z(x,y)/dn$ (in units of $\mu_B$/electron or equivalently $\mu_B$/state) (see Methods). Figure 3a shows that at low filling, $\nu = 0.737$, isolated patches of positive (paramagnetic-like) $m_z(x,y,\nu_\uparrow)$ are present along with areas of negative (diamagnetic-like) response. Equivalently, $B_z^{ac}(x,y)$ can be inverted into current density $J^{ac}(x,y)$ (in units of µA/µm) as shown in Fig. 3b and in the Supplementary Video 3. Note that magnetization $M$ in 2D (magnetic moment per unit area) is given in units of current (A), which for the case of a uniformly magnetized domain describes the persistent current that circulates along the edges of the domain. Figure 3b shows these sharp current channels that circulate along the patch edges.

**Tomography**

Upon increasing $\nu$, $m_z(x,y,\nu_\uparrow)$ evolves into a complex pattern of paramagnetic-like and diamagnetic-like domains as shown in Fig. 3d for $\nu = 0.966$. The corresponding $m_z(x,y,\nu_\downarrow)$ and $\Delta m_z(x,y,\nu)$ are presented in Figs. 3e,f. To gain insight into the underlying mechanism, we construct a tomographic presentation of the full data sets of $m_z(x,y,\nu_\uparrow)$, as depicted in Fig 3c, which shows how the local $m_z$ evolves as a function of $\nu$ (see Supplementary Videos 3 and 4 for the full evolution of $m_z(x,y,\nu_\uparrow)$, $m_z(x,y,\nu_\downarrow)$, $\Delta m_z(x,y,\nu)$, and $J^{ac}(x,y,\nu_\uparrow)$ with $\nu$).

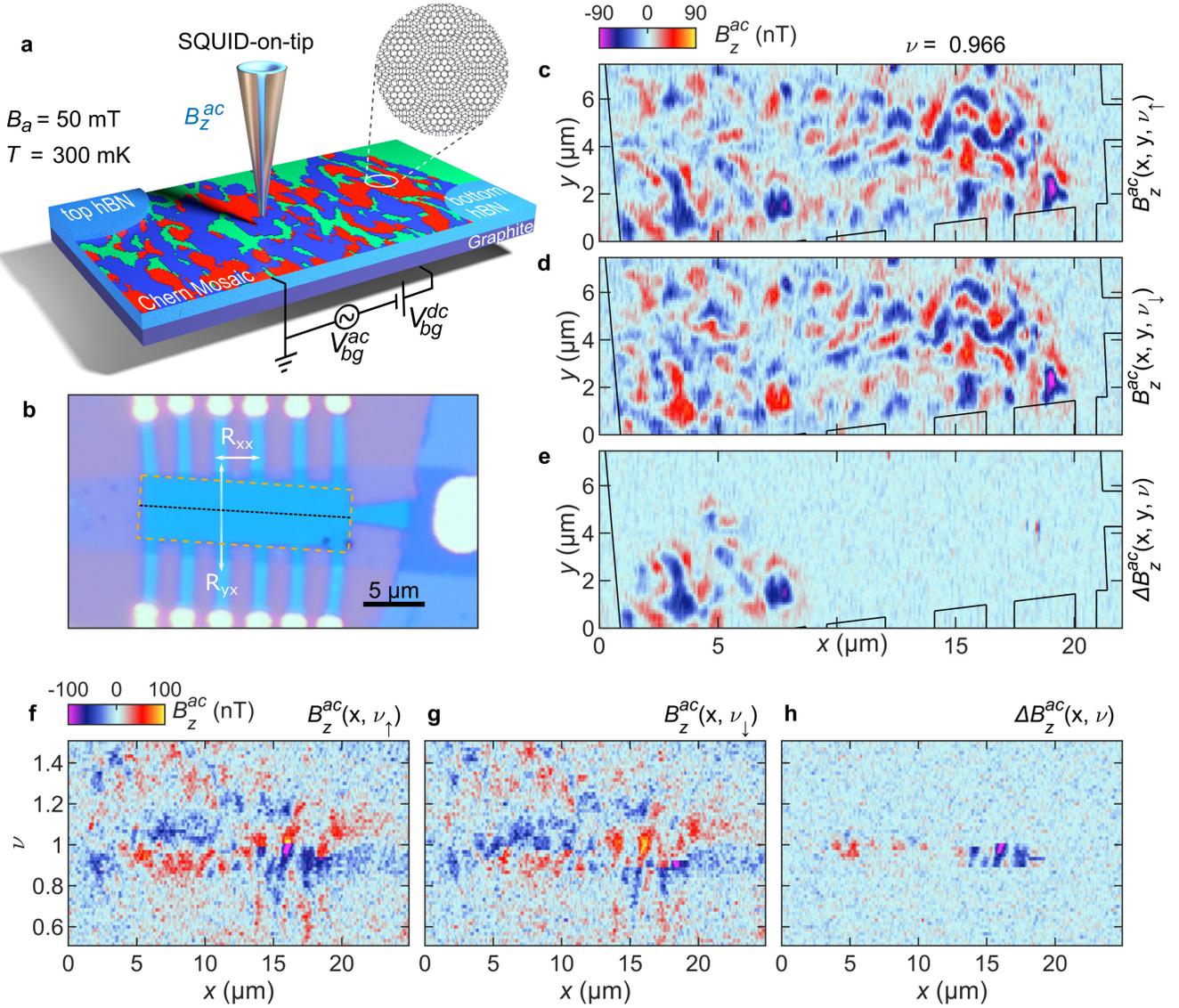

**Fig. 2. Imaging the local magnetic response. a**, Schematic layout showing backgate voltage $V_{bg}^{dc} + V_{bg}^{ac}$ applied to the MATBG sample and the corresponding change in the local magnetic field $B_z^{ac}(x, y)$ is imaged using the scanning SOT. The Chern mosaic is shown schematically in the MATBG. **b**, Optical image of the sample patterned into a Hall bar geometry with the scanning window indicated by the orange rectangle. The contacts used for measurements of $R_{xx}$ and $R_{yx}$ are indicated. **c**, $B_z^{ac}(x, y, \nu_\uparrow)$ at $\nu = 0.966$ upon sweeping $\nu$ up, $\nu_\uparrow$. See Supplementary Video 2 for a full range of $\nu$. The black contour outlines the sample edges. **d**, Same as **c** upon sweeping $\nu$ down, $B_z^{ac}(x, y, \nu_\downarrow)$. **e**, The hysteresis image attained by numerical subtraction of (**d**) from (**c**), $\Delta B_z^{ac}(x, y, \nu) = (B_z^{ac}(x, y, \nu_\uparrow) - B_z^{ac}(x, y, \nu_\downarrow))/2$. The rms value of $\Delta B_z^{ac}$ variations in the non-hysteretic part of (**e**) is 4.5 nT providing a measure of the error bar of the $B_z^{ac}$ data. **f**, $B_z^{ac}(x, \nu_\uparrow)$ measured along the black dotted line in (**b**) vs. $\nu$ upon sweeping $\nu$ up. **g**, Same as (**f**) upon sweeping $\nu$ down, $B_z^{ac}(x, \nu_\downarrow)$. **h**, The hysteresis image $\Delta B_z^{ac}(x, \nu) = (B_z^{ac}(x, \nu_\uparrow) - B_z^{ac}(x, \nu_\downarrow))/2$.

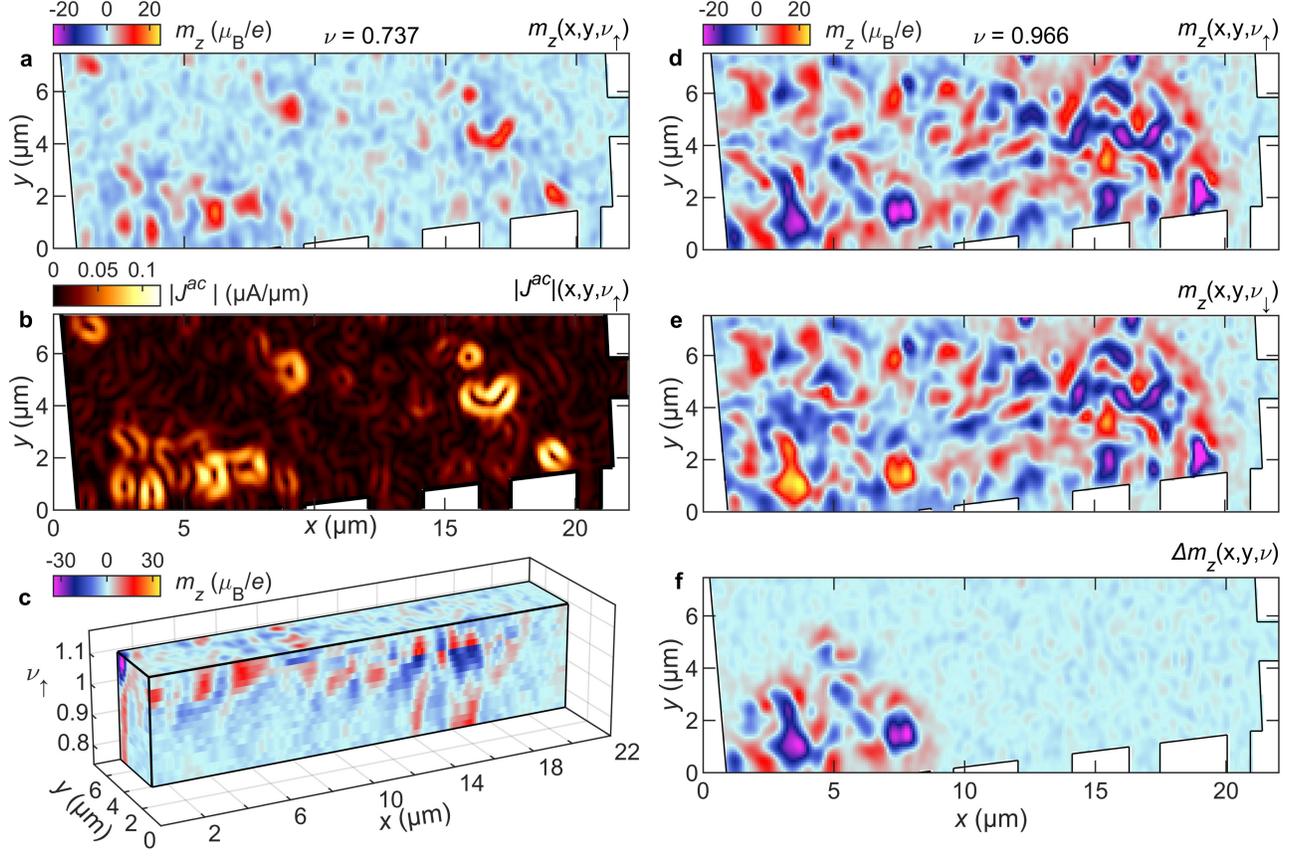

**Fig. 3**. **Imaging local magnetization and equilibrium currents. a**, Differential magnetization $m_z(x,y) = dM_z(x,y)/dn$ reconstructed from $B_z^{ac}(x,y)$ induced by filling factor modulation of $\nu_{ac} = 0.083$ at $\nu_\uparrow = 0.737$. The white areas are outside the sample. **b**, Equivalent description of magnetization in terms of equilibrium currents, $\bm{J} = \nabla \times \bm{M}_z$. Shown is the magnitude of equilibrium current density $|\bm{J}^{ac}(x,y)|$ reconstructed from $B_z^{ac}(x,y)$ (see Supplementary Video 3 for $\bm{J}^{ac}(x,y)$ for a range of $\nu$). **c**, Tomographic view of $m_z(x,y,\nu_\uparrow)$. **d-f**, $m_z(x,y,\nu_\uparrow)$ (**d**), $m_z(x,y,\nu_\downarrow)$ (**e**), and the magnetization hysteresis $\Delta m_z(x,y,\nu) = (m_z(x,y,\nu_\uparrow) - m_z(x,y,\nu_\downarrow))/2$ (**f**) at $\nu = 0.966$ (see Supplementary Video 4 for the tomographic view over the full range of $\nu$). The red (blue) colors indicate paramagnetic-like (diamagnetic-like) local differential magnetization. The rms value of $\Delta m_z$ variations in the non-hysteretic part of (f) is 0.69 $\mu_B/e$, providing a measure of the error bar of the $m_z$ data.

The tomographic data set reveals a number of key observations: (*i*) The measured local differential magnetization $m_z$ is very large, reaching values of 25 $\mu_B$ per electron, which is consistent with theoretical predictions [27] and with our calculations (Methods), and provides strong evidence for the orbital origin of the magnetization. (*ii*) Observation of the orbital magnetism establishes that the valley degeneracy is lifted and the time reversal symmetry is broken. The valley polarization commences well below $\nu = 1$. (*iii*) The pronounced variations in the equilibrium local magnetization point out the local variations in the band structure and in the Berry curvature. (*iv*) Some regions of the sample show reversible magnetization with $\nu$ whereas others are hysteretic. (*v*) The hysteretic and non-hysteretic regions have comparable $m_z$, suggesting that they are controlled by the same mechanism. (*vi*) $m_z(x,y)$ displays coexistence of paramagnetic-like (red) and diamagnetic-like (blue) patches. Since a paramagnetic dipole moment, $\bm{m}$, is favorable in terms of potential energy $U_m = -\bm{m} \cdot \bm{B}$, the presence of diamagnetic blue patches is surprising. (*vii*) At no filling factor the entire sample is in a single paramagnetic or diamagnetic-like state. (*viii*) In conventional ferromagnets, unfavorable magnetic domains are present to minimize the $B^2$ energy term and their size shrinks in jump-wise manner upon increasing $B_a$, giving rise to Barkhausen noise. One would therefore expect to see blue patches flipping abruptly to red. Remarkably, $\Delta m_z(x,y)$ in Fig. 3f shows that the hysteretic jumps involve intricate flipping of both favorable and unfavorable domains.

**Origin of orbital magnetization**

To analyze the results further, one needs to understand the origin of magnetization in MATBG. The orbital magnetization arises from the Berry curvature and the topological nature of the bands, and has two contributions [25,27,35–37]. The first is the self-rotation of the wave packet, $M_{SR}$, and the second is the electric-field-induced transverse drift velocity of the wave packet. We refer to the latter as Chern magnetization, $M_C$, since it acquires its main contribution in the topological Chern gaps [35]. The two contributions have analogs in the quantum Hall effect, where $M_{SR}$ is similar to the magnetization due to the cyclotron motion of the bulk electrons, and $M_C$ is equivalent to the magnetization induced by the equilibrium (ground-state) topological currents flowing in the edge states [39,40]. It follows then that $M_C$ is generally of opposite sign to $M_{SR}$ (see further explanation in Methods).

The magnitudes and evolutions of $M_{SR}$ and $M_C$ in the metallic state are not universal and depend on the details of the band structure. To evaluate the magnetizations we carry out single particle band structure calculations (Methods) [27]. We assume that at CNP the system is degenerate with $C = 0$, and that the flavor degeneracy is fully lifted spontaneously upon increase of $\nu$. Figure 4a shows the calculated band structure of a single valley-polarized ($K$) flat band with the corresponding self-rotation magnetization presented in Fig. 4b. The resulting integrated $M_{SR}$ (Fig. 4c) grows gradually with $\nu$, shows an upturn upon approaching the top of the band, and reaches values of $M_{SR} \cong 1$ to 3 $\mu_B$ per unit cell (u.c.) at $\nu = 1$ (depending on band structure parameters, see Extended Data Fig. 4). Since there are no states in the gap, $M_{SR}$ remains constant until the next flat band with opposite valley polarization ($K'$) starts to be occupied (Methods). Since $K$ and $K'$ polarized flat bands have $M_{SR}$ of opposite sign, the integrated $M_{SR}$ decreases for $\nu > 1$. The corresponding $m_{SR} = dM_{SR}/dn$ (Fig. 4d), which is the self-rotation component of the experimentally measured total $m_z$, is quite small at low fillings and increases rapidly on approaching $\nu = 1$, finally reaching values of about 20 $\mu_B$ per electron. The $m_{SR}$ vanishes in the gap and has an opposite sign in the following band. The $M_C$ has a behavior that is quite opposite to $M_{SR}$ (Fig. 4c). It is mostly very small in the compressible state and attains its maximal (negative) value of $M_C^{max} = C\Delta e/h$ when the chemical potential $\mu$ reaches the top of the Chern insulator gap ($\Delta$ is the gap energy, $e$ is the elementary charge, $h$ is Planck's constant, and $C = -1$ in Fig. 4c). Note that in the metallic state the orbital magnetism depends on the evolution of the Berry curvature with the chemical potential. In contrast, when $\mu$ resides in the gap, $M_C$ and $m_C$ have a universal behavior, $m_C = dM_C/d\mu = Ce/h$, because the Berry-curvature flux of a full band is quantized and is equal to $C$, leading to a finite $M_C$ and $m_C$ if and only if $C$ is nontrivial [25]. Thus, measurement of a finite $M_C$ and $m_C$ constitutes a direct observation of a topological gap with the sign of $C$ given by the sign of $M_C$ and $m_C$.

**Magnetization reversal**

The $K$ and $K'$ polarized flat bands are degenerate at zero field. At low $\nu$, a small $B_a$ favors the valley polarization (e.g. $K$) with positive total magnetization, $M_z = M_{SR} + M_C \cong M_{SR}$. As the local chemical potential $\mu$ enters the gap, the $M_C$ contribution dominates and may cause $M_z$ to change sign, as indicated by the red dot in Fig. 4c. When this happens, it is energetically favorable to re-condense all states through a first-order phase transition to the opposite valley ($K'$) in order to realign $M_z$ with $B_a$. This leads to a discontinuous flip in the polarity of $M_z$ and $m_z$ and inversion of all the curves in Figs. 4c,d [15,32,35], as depicted in Extended Data Fig. 5a.

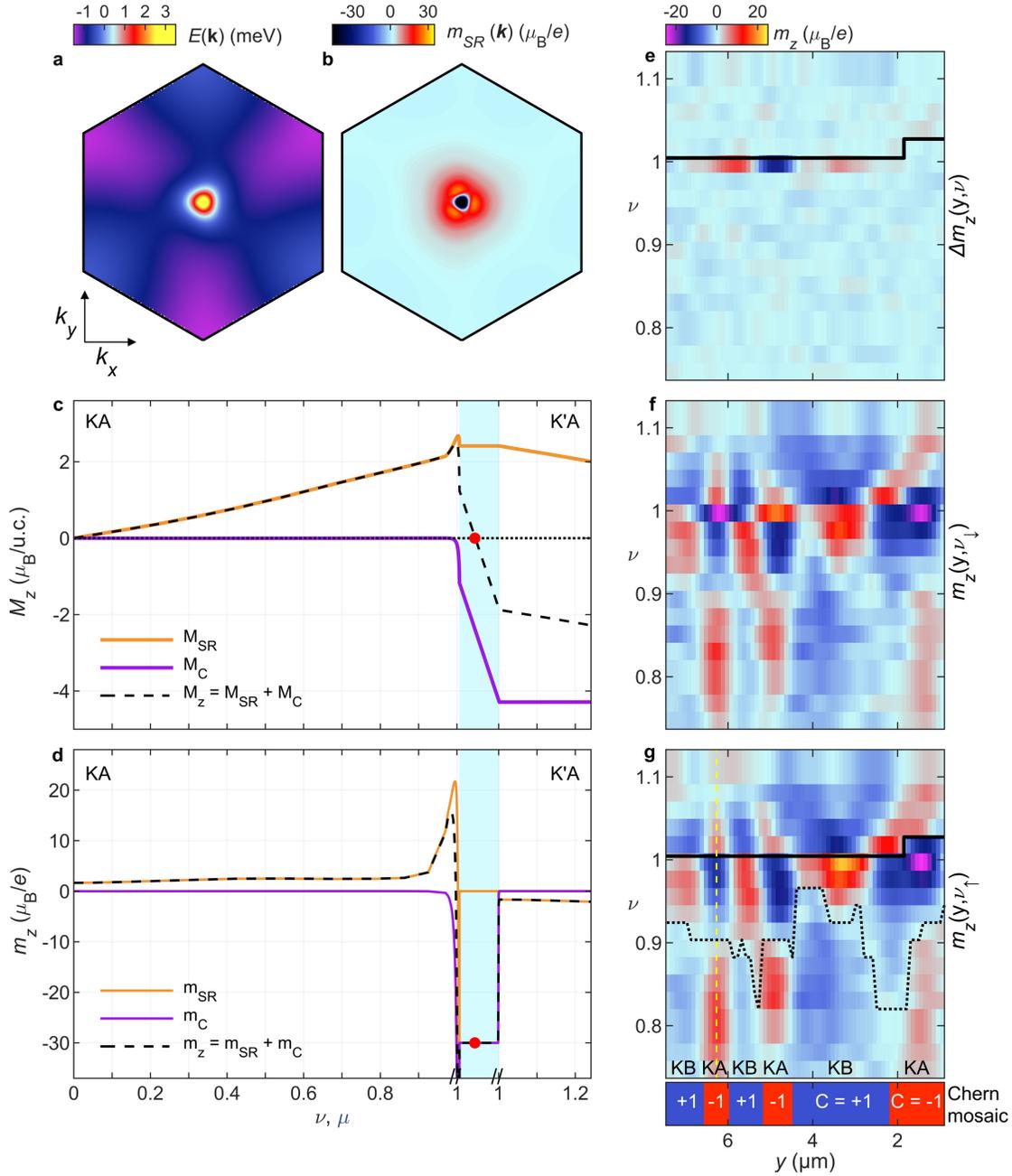

**Fig. 4. Evolution of orbital magnetism. a**, Single particle band structure of the $KA$ polarized flat band in the first mini-Brillouin zone for twist angle $\theta = 1.08°$, staggered potential $\delta = 17$ meV, and tunneling ratio $w = 0.95$ (see Methods). **b**, Momentum-resolved self-rotation magnetization $m_{SR}(\mathbf{k})$ in the $KA$ band. **c**, Calculated evolution of the momentum-integrated magnetization $M_z$ and its two components $M_{SR}$ and $M_C$ vs. the filling factor $\nu$ in the compressible states or vs. the chemical potential $\mu$ in the Chern gap (cyan), which is taken to be $\Delta = M_C h/Ce = 7$ meV as an example. Starting from zero magnetization at CNP, the $M_{SR}$ (orange) grows positive in the $KA$ band, remains constant in the gap, and then decreases in the following $K'A$ flat band. $M_C$ (purple) is very low in the compressible state and grows negative close to the band edge, followed by linear increase with $\mu$ in the Chern gap, $dM_C/d\mu = Ce/h$ with $C = -1$. The total $M_z$ (dashed) changes sign in the gap region (red dot), above which abrupt hysteretic flipping of the polarity of $M_{SR}$ and $M_C$ may occur due to electron recondensation from the $K$ to $K'$ valley. Such flipping is not shown here for clarity and is presented

in Extended Data Fig. 5. Although theoretically the transition through the gap should be sharp due to zero density of states, in practice the presence of disorder broadens significantly the gap region. **d**, The corresponding evolution of the differential magnetization $m_z$ and its $m_{SR}$ and $m_C$ components. See Extended Data Fig. 5 for details in the vicinity of $\nu = 1$. **e-g**, Tomographic slices of $\Delta m_z(y, \nu)$ (**e**), $m_z(y, \nu_\downarrow)$ (**f**), and $m_z(y, \nu_\uparrow)$ (**g**) across the sample at $x = 15.2$ µm. In (g) the solid black line marks the first-order phase transition $\nu_{\uparrow f}(y)$ where $m_z(y, \nu_\uparrow)$ flips discontinuously due to electron recondensation from $K$ to $K'$ valley, coinciding with the top of the hysteretic signal $\Delta m_z(y, \nu)$ in (e). The dotted black line marks $\nu_o(y)$ along which $m_z$ vanishes, demarcating the transition from $m_{SR}$ to $m_C$ dominated regions. A line cut along the dotted yellow dashed line is presented in Extended Data Fig. 5b. The bottom strip shows schematically the Chern mosaic with $C = 1$ (blue) and $C = -1$ (red) patches and their valley and sublattice polarizations.

The tomographic slice of the measured $m_z(y, \nu_\uparrow) = dM_z(y, \nu_\uparrow)/dn$ in Fig. 4g shows this flipping mechanism at play. Focusing on $C = -1$ patches (denoted at the bottom of Fig. 4g), $m_z$ starts off red (positive) and intensifies with $\nu$, reflecting the contribution of $m_{SR}$, reaching peak values of ~10 $\mu_B/e$, consistent with the calculations in Fig. 4d. The magnetization $m_z$ then drops rapidly (but reversibly) through $m_z = 0$ (demarcated by the dotted line, $\nu_o(y)$) to intense blue due to the onset of the negative $m_C$ contribution in the Chern gap. Upon further filling, a discontinuous flip from blue to red occurs (marked by the solid black line, $\nu_{\uparrow f}(y)$), where the system recondenses all the carriers to the opposite valley through a first-order phase transition. Depending on the height of the barrier between these two states, the process can either be hysteretic or non-hysteretic as seen in Fig. 4e and discussed below.

The condition for the flip to occur is $-M_C^{max} > M_{SR}^{max}$, where $M_{SR}^{max}$ is the value of $M_{SR}$ at the top of the band. By integrating the $m_z$ tomographic data over $\nu$ up to $\nu_o(x, y)$ we attain a map of $M_{SR}^{max}(x, y)$ (Fig. 5b). The $M_{SR}^{max}$ in the red patches attains typical values of 1 to 2 µB/u.c., consistent with calculations in Fig. 4c. By integrating $m_z(x, y)$ from $\nu_o(x, y)$ to the filling factor at which the flipping occurs, $\nu_{\uparrow f}(x, y)$, we attain a map of the lower bound of $M_C^{max}$ (see Methods) as shown in Fig. 5c. The patches that are red in Fig. 5b appear as dark blue/purple in Fig. 5c with values of $M_C^{max}$ reaching down to −4 µB/u.c., thus explaining the mechanism of the magnetization flipping. Moreover, the finding of the negative local $M_C$ magnetization giving rise to the magnetization flipping at $\nu = 1$ constitutes a direct local observation of the presence of a topological gap with a negative Chern number $C = -1$, which provides a strong evidence that at least locally the sample is in fact a zero-field Chern insulator [35]. From the relation $M_C^{max} = C\Delta e/h$ we can evaluate the lower bound of the local Chern gap, which for $M_C^{max} = -4$ µB/u.c. is $\Delta = 7$ meV, consistent with recent compressibility studies [29].

**Chern mosaic**

Figure 4g clearly shows that together with the paramagnetic-like "red" patches ($C = -1$) that behave as discussed above, there are numerous diamagnetic-like "blue" patches ($C = 1$). These patches display an opposite sequence starting from light blue $m_z$, crossing through zero to dark red, and flipping abruptly to dark blue. The presence of these patches with negative $M_{SR}$ is highly surprising since they appear to be exceedingly unfavorable energetically. Even more puzzling is that according to the mechanism described above, one would expect the hysteretic jumps to reflect only abrupt flipping of negative $m_z$ to the energetically favorable positive $m_z$. Remarkably, Figs. 4g and 3f show that the hysteretic jumps involve flipping of both blue and red patches.

The origin of the diamagnetic-like (blue) patches cannot be attributed to an antiferromagnetic desire to minimize magnetic energy. The measured stray magnetic field $B_{stray}$ from neighboring red patches is of the order of µT while $B_a = 50$ mT, such that the energy cost associated with a blue patch of size $S$ ($\approx SM_{SR}B_a$) is some four orders of magnitude larger than the antiferromagnetic energy gain ($\approx SM_{SR}B_{stray}$). Thus, the origin of these patches must be found elsewhere.

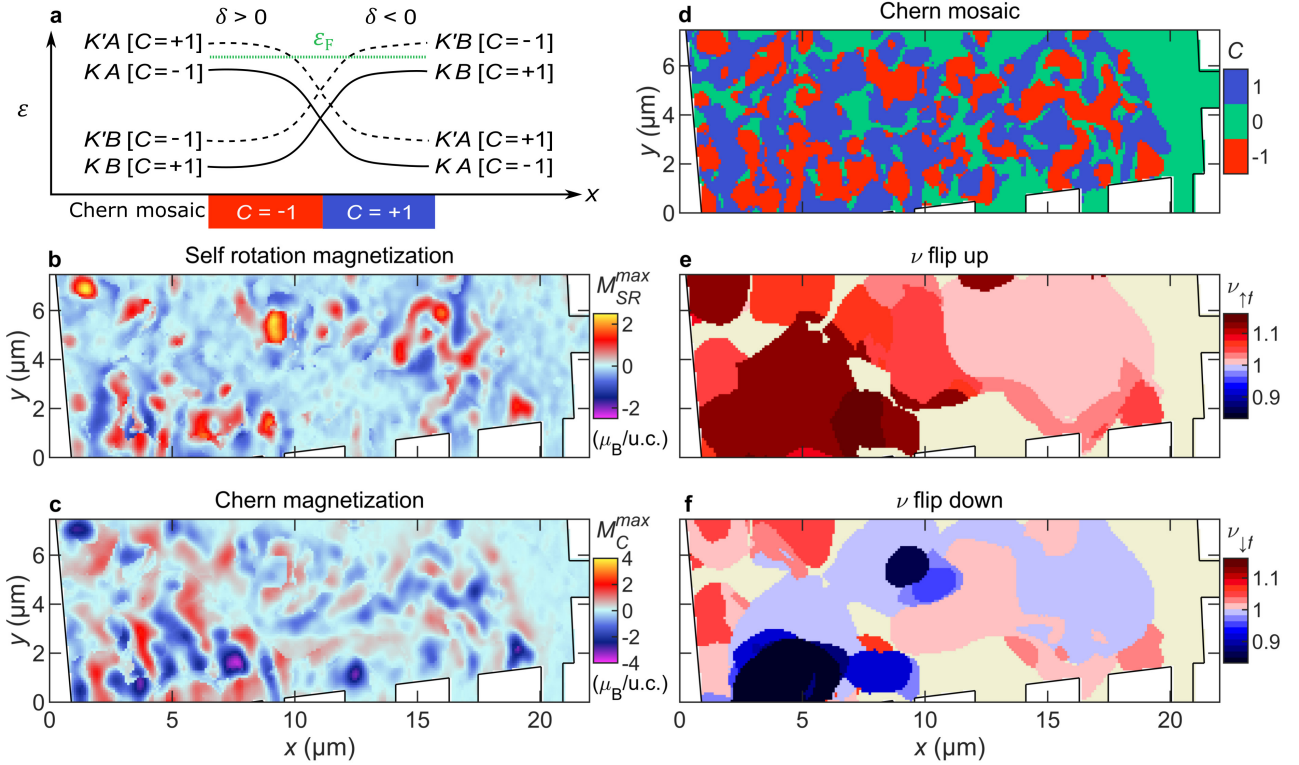

**Fig. 5**. **Chern mosaic.** **a**, Schematic diagram of the flat bands in neighboring $C = -1$ and $C = 1$ patches imposed by the sign of the staggered potential $\delta$ at $\nu = 1$ and neglecting spin. The bands below the Fermi energy $\varepsilon_F$ are occupied. **b**, The self-rotation magnetization $M_{SR}^{max}(x,y)$ derived by integration of the measured $m_z(x,y,\nu_\uparrow)$ up to $\nu_o(x,y)$. Regions outside the sample are white. **c**, Chern magnetization $M_C^{max}(x,y)$ derived by integration of $m_z(x,y,\nu_\uparrow)$ from $\nu_o(x,y)$ up to $\nu_{\uparrow f}(x,y)$. **d**, Chern mosaic map derived from the evolution of $m_z(x,y,\nu_\uparrow)$ showing the $C = 1$ ($KB$ polarization, blue), $C = -1$ ($KA$, red), and $C = 0$ or semimetallic intermediate regions (green). **e-f**, Maps of the local filling factor at which the flipping of $m_z(x,y)$ occurs upon sweeping $\nu$ up, $\nu_{\uparrow f}(x,y)$ (**e**), and down, $\nu_{\downarrow f}(x,y)$ (**f**) in steps of $\Delta\nu = 0.02$. In the yellow-grey areas no discontinuous flipping of $m_z$ is resolved.

Since $M_C = C\Delta e/h$, the detection of patches with opposite signs of $M_C$ in the vicinity of $\nu = 1$ (Fig. 4g) is a direct local observation of a mosaic of positive and negative Chern numbers. The Chern number is determined by the product of valley ($K,K'$) and sublattice ($A,B$) polarizations, of which there are four possible competing combinations. If we denote a red patch (with $C = -1$) as $KA$ polarized, there are three polarization possibilities to consider for the neighboring patches: $K'A$, $KB$, or $K'B$ [22]. Since $K'B$ polarization also results in $C = -1$, it can appear in other red patches, but cannot be present in blue patches. Thus, the observed blue patches must be polarized to either $K'A$ or $KB$. Both possibilities have a magnetic energy cost ($SM_{SR}B_a$) and an additional cost of either $K$-$K'$ or $A$-$B$ domain wall energy, which is proportional to the length of the interface. The $KA$-$K'A$ mosaic is unlikely since there is no obvious external mechanism that can enforce it. In contrast, the $KA$-$KB$ mosaic can arise naturally by sublattice polarization, which breaks $C_2$ symmetry that is associated with gap opening. Such gap opening must be present in any case to form the observed zero-field Chern insulator state.

We propose two possible mechanisms that can explain such a space-dependent pattern of breaking of sublattice symmetry, which is implied by our measurements. The first mechanism is based on Hartree-Fock calculations, which predict that the system may break the sublattice symmetry spontaneously near $\nu = 1$, even in the absence of an external sublattice potential [41,42]. If the hBN substrate potential is incommensurate with the graphene moiré lattice, it acts as a spatially varying field that couples to the local sublattice polarization. In the presence of strain and angle disorder, this field becomes effectively random. In this case, according to the Imry-Ma argument [43], the system is expected to break up into domains of opposite sublattice polarization, whose size is large in the limit of small random field, thus forming a spontaneous Chern mosaic stabilized by disorder.

Alternatively, if the twist angles between the two graphene layers and between the hBN layer and the adjacent graphene layer create nearly commensurate moiré lattices (see Methods and Extended Data Fig. 6), it has recently been shown [20,21,44] that in absence of disorder a position-dependent periodic sublattice polarization may occur, giving rise to semimetallic or gapped Chern bands with $C$ equal either to 1, −1, or 0. Combining these predictions with the presence of strain [27], structural relaxation [45], and twist-angle disorder [40] in MATBG, we envision a non-homogenous non-periodic Chern mosaic, where $KA$ polarized patches with positive $M_{SR}$ and $C = -1$ and $KB$ polarized patches with negative $M_{SR}$ and $C = 1$, are energetically stabilized by the combination of disorder and a coarse-grained substrate potential. Figure 5a shows a schematic of such a $KA$-$KB$ interface at $\nu = 1$ (neglecting spin) where the bands invert across the domain wall and the Chern number changes sign.

By analyzing the $m_z(x, y, \nu_\uparrow)$ signal at filling factors below $\nu_0(x, y)$ (Methods), we can reconstruct the Chern mosaic structure of our sample as shown in Fig. 5d. We find isolated $C = -1$ patches (red) of characteristic size of a few µm that are embedded in $C = 1$ (blue) and $C = 0$ or semimetallic regions (green). Note also that none of the phases percolates across the sample, which explains the observation of the AHE in the transport measurements in Figs. 1c-g without full quantization, as proposed in [20].

**Dynamics of the magnetization reversal**

In the derived picture, the $A$ and $B$ sublattice pattern is imprinted in the sample by the local variations in the lattice structure. In the presence of a small positive $B_a$ and $\nu < 1$, the red $KA$ patches with positive $M_{SR}$ are favorable energetically, while the blue $KB$ patches with negative $M_{SR}$ are unfavorable. This raises the question of why their valley polarization does not flip to $K'B$ thus rendering all the patches the favorable (red) magnetization direction. The answer lies in the fact that the $U_m = -\boldsymbol{m} \cdot \boldsymbol{B_a}$ cost of the $KB$ patches is balanced by avoidance of the formation of $K$-$K'$ domain walls [22] with line energy which is estimated to be $U_{dw} \approx 0.05$ meV/nm [22,46] (Methods). Equating $U_m = \pi D^2 M_{SR} B_a/4 = \pi D U_{dw}$ for a patch of diameter $D$ and using $M_{SR} = 1\ \mu_B$/u.c., we attain $D_{min} = \frac{4U_{dw}}{M_{SR}B_a} \approx 5$ µm. This means that if the diameter of the $KB$ patch is smaller than $D_{min}$, it is energetically favorable to keep it in the $K$ valley and pay the cost of negative $M_{SR}$, rather than forming a $K$-$K'$ domain wall.

Since the typical size of the Chern patches in Fig. 5d is smaller than the estimated $D_{min}$, at low $B_a$ and $\nu < 1$ the entire sample is apparently in a single valley polarized state $K$, which comprises a Chern mosaic of patches of different $C$. Upon entering the Chern gap, the situation changes due to the development of large $M_C$ that is of opposite polarity to $M_{SR}$ (Fig. 4g). In this situation, the $KA$ patches become energetically unfavorable (blue), and should thus recondense abruptly into $K'A$ state [35], as discussed above. The individual patches, however, cannot flip their valley polarization independently because of the high $K$-$K'$ domain wall energy, $U_{dw}$. As a result, the hysteretic jumps must occur collectively in which a large domain, comprising of a number of $C = -1$ and $C = 1$ patches, reverses its valley polarization as a whole. Note that such a large domain has a significantly smaller average $M_Z$ leading to even larger estimated minimal domain size $D_{min}$. Note also that the $KB$ patches, which at low $\nu$ were energetically unfavorable because of negative $M_{SR}$, become energetically favorable (red) in the gap region due to positive $M_C$, and therefore their flipping back to energetically unfavorable state (blue) cannot be explained without such a collective effect. Figure 4g provides direct evidence for such an abrupt collective $K$ to $K'$ valley inversion where the opposite color patches flip their magnetization simultaneously across the horizontal section of the black $\nu_{\uparrow f}(y)$ line. Importantly, once a domain has flipped its valley polarization, a $K$-$K'$ domain wall is formed along its edges providing the means for controllable creation of valley domain walls by backgate voltage.

Figure 5e presents the map of the filling factor, $\nu_{\uparrow f}(x,y)$, at which the flipping of the local magnetization occurs upon sweeping $\nu_\uparrow$ (Methods). It reveals that the magnetization reversal process mainly involves flipping of large domains on a scale of about 5 µm as well as a number of smaller domains. Also, there is a significant spread in the $\nu_{\uparrow f}$ values ranging from 1.01 in the right-hand side of the sample to 1.17 on the left side. Some areas of the sample (yellow-grey) show no detectable flipping. On sweeping $\nu_\downarrow$, the $\nu_{\downarrow f}(x,y)$ map in Fig. 5f displays a similar behavior with values ranging from 1.07 down to 0.84, albeit with somewhat smaller typical domain sizes. Additionally, the left part of the sample shows a substantially larger hysteresis in $\nu_f$, in good correlation with the larger $M_C^{max}$ in Fig. 5c in this area. These observations can be related to the fact that the twist angle $\theta$ is slightly higher in the left part of the sample as show in Extended Data Fig. 7. From $\Delta m_z(x,y,\nu)$ one can extract the evolution of the magnetization hysteresis as a function of $\nu$ (Extended Data Fig. 8), which is found to be in good agreement with transport hysteresis in Fig. 1d, and explains the progressive evolution of the hysteresis in the minor loops in Fig. 1g.

With increasing $B_a$, $D_{min}$ should decrease and eventually drop below the Chern mosaic patch size. In this case, the magnetic energy dominates over $U_{dw}$ and each patch can act independently with no collective reversals similar to the case of magnetically doped topological insulators [47]. As a result, we anticipate that at $\nu < 1$ all the patches will align with positive $M_{SR}$, forming $KA$ and $K'B$ patches separated by $K$–$K'$ domain walls, leading to the suppression of the hysteresis with $B_a$, consistent with the transport data in Figs. 1c-f. This regime will be a subject of future studies.

Our finding of the Chern mosaic in a sample that was not aligned intentionally with hBN suggests that this new type of topological disorder can be ubiquitous in MATBG devices (see Methods and Extended Data Fig. 6 for further discussion). Moreover, identifying magnetism through transport measurements relies predominantly on observation of hysteresis [13–15,19,30,32,48], which requires a first-order transition. Since our local measurements reveal Berry curvature magnetism and Chern gaps even in the absence of local hysteresis, orbital magnetism in MATBG could be more omnipresent than reflected by transport. Being imprinted by the substrate potential, it dictates the local flavor polarization, providing a unique mechanism for manipulation and possible utilization of flavor-dependent electronic properties, opening the route for "flavortronic" devices. In particular, the valley and Chern domain walls can host novel edge states [49]. Domain walls between regions of opposite sublattice polarization in the Chern mosaic are expected to carry two co-propagating chiral edge states, while domain walls where both the valley and the sublattice polarizations flip may carry counter-propagating valley-helical modes of opposite valley polarization. At such a domain wall, the system is likely to favor an inter-valley coherent state, where the charge and valley currents are carried by a gapless collective mode [22]. Studying the structure of these domain walls and their transport properties is an interesting direction for future investigations.

**Acknowledgments** The authors thank M. E. Huber for SOT readout system. This work was supported by the European Research Council (ERC) under the European Union's Horizon 2020 research and innovation program (grant No 785971), and by the Israel Science Foundation ISF (grants No 921/18 and 994/19). A.S. and E.B. acknowledge support from the Israel Science Foundation's Quantum Science and Technology grant no. 2074/19 and from CRC 183 of the Deutsche Forschungsgemeinschaft (Project C02). E.B. was supported by the European Research Council (ERC) under the European Union's Horizon 2020 research and innovation programme (grant No 817799). E.Z. acknowledges the support of the Andre Deloro Prize for Scientific Research and Leona M. and Harry B. Helmsley Charitable Trust grant #2112-04911. D.K.E. acknowledges support from the Ministry of Economy and Competitiveness of Spain through the "Severo Ochoa" program for Centres of Excellence in R&D (SE5-0522), Fundació Privada Cellex, Fundació Privada Mir-Puig, the Generalitat de Catalunya through the CERCA program, funding from the European Research Council (ERC) under the European Union's Horizon 2020 research and innovation programme (grant No. 852927). P.S. acknowledges support from the European Union's Horizon 2020 research and innovation programme under the Marie Skłodowska-Curie grant agreement No. 754510. G.D.B. acknowledges support from the "Presidencia de la Agencia Estatal de Investigación" (Ref. PRE2019-088487). K.W. and T.T. acknowledge support from the Elemental Strategy Initiative conducted by the MEXT, Japan (Grant Number JPMXP0112101001) and JSPS KAKENHI (Grant Numbers 19H05790, 20H00354 and 21H05233).

**Author contributions** S.G., M.B. and E.Z. designed the experiment. S.G. and M.B. performed the measurements. M.B., A.U. and S.G. performed the analysis. P.S., G.B. and D.K.E. designed and fabricated the sample and contributed to the analysis of the results. I.R. fabricated the SOTs and Y.M. fabricated the tuning forks. J.X. and B.Y. performed the band structure calculations. A.Y.M. developed the magnetization reconstruction code. E.B, A.S., and K.P. contributed to the analysis and the theoretical modeling. K.W. and T.T. provided the hBN crystals. M.B., A.U., S.G. and E.Z. wrote the manuscript. All authors participated in discussions and in writing of the manuscript.

**Competing interests** The authors declare no competing interests.

**Data availability** The data that support the findings of this study are available from the corresponding authors on reasonable request.

**Code availability** The band structure calculations and magnetization reconstruction codes used in this study are available from the corresponding authors on reasonable request.

## Methods

**Device fabrication**

The MATBG device was fabricated using a cut and stack technique. All flakes were first exfoliated on a Si/SiO$_2$ (285 nm) substrate and later picked up using a polycarbonate (PC)/polydimethylsiloxane (PDMS) stamp. All the layers were picked up at a temperature of ~100 °C. The graphene was initially cut with an AFM tip, to avoid strain during the pick-up process. The PC/PDMS stamp was used to pick up the top hBN and then the first graphene layer. Before picking up the second graphene layer, the stage was rotated by an angle of 1.1° to 1.2°. Finally, the bottom hBN and bottom graphite gate were picked up. The finalized stack was dropped on a Si/SiO$_2$ substrate by melting the PC at 180 °C and etched into a Hall bar using CHF$_3$/O$_2$. The 1D contacts were formed by evaporating Cr (5 nm)/Au (50 nm). The optical and AFM images of the different fabrication steps are shown in Extended Data Figs. 1a-d. All the presented experimental data were acquired on this device which was originally studied in Ref. [19]. The device has subsequently undergone several additional thermal cycles before the current study. A high resolution AFM image of the device acquired after the present study is presented in Extended Data Fig. 1e. The dark spots in the image are due to damage caused to the device after the completion of the measurements presented in this paper. We do not observe any clear correlation between topographic structure and the measured Chern mosaic.

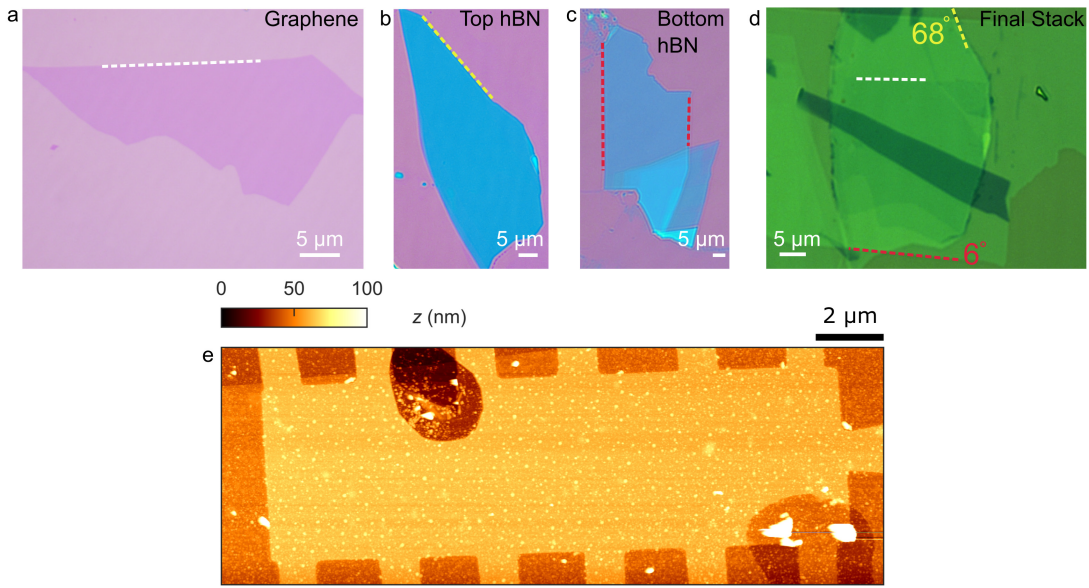

**Extended Data Fig. 1**. **Optical and AFM images of the device. a-d**, Optical images of graphene (**a**), top hBN (**b**), bottom hBN (**c**), and the final stack (**d**). The white (graphene), yellow (top hBN), and red (bottom hBN) dashed lines show that the top (bottom) hBN layer is at an angle of 68° (6°) with respect to the graphene. **e,** Zoomed-in AFM image of the device. The two large dark spots are due to damage caused to the device after the measurements presented in this paper.

There was no intentional alignment between the graphene bilayer and either of the hBN flakes. The white, yellow and red dashed lines in Extended Data Figs. 1a-c mark the naturally broken edges of the graphene, top hBN, and bottom hBN flakes respectively. Extended Data Fig. 1d shows that the twist angle $\theta_{GBN}$ between the graphene and the bottom hBN layer was 6°, while the relative twist angle with the top hBN was 68°, which is equivalent to 8° under the symmetries of the system. These twist angles are within the range of possible angles that can create commensurability between the graphene-graphene and the graphene-hBN moiré lattices as shown in Extended Data Fig. 6.

**Transport measurements**

Four-point transport measurements were performed at $T = 300$ mK using standard lock-in techniques with a bias current of $I = 10$ nA rms at 11 Hz. Figures 1a,b show the longitudinal and transverse resistances $R_{xx}$ and $R_{yx}$. The $R_{xx}$ data is reproduced in Extended Data Fig. 2a along with the map of the Landau levels in Extended Data Fig. 2b. We extract the following parameters: back gate capacitance $C_{bg} = 450$ nF·cm$^{-2}$ = $2.81 \times 10^{12}$ $e$ cm$^{-2}$V$^{-1}$; $V_{bg}$ corresponding to $\nu = \pm 4$ is 0.948 V and $-0.971$ V respectively, and the carrier density for full filling of the flat band $n_s = 2.69 \times 10^{12}$ cm$^{-2}$, which corresponds to a twist angle of $\theta \approx \sqrt{\frac{\sqrt{3}}{8} a^2 n_s} = 1.08°$ with $a = 0.246$ nm.

At low filling factors the total orbital magnetization $M_z = M_{SR} + M_C$ is dictated by $M_{SR}$. Since $K$ and $K'$ polarized bands are equivalent, a small applied field will select the valley ($K$) with positive $M_{SR}$ which has a negative Chern gap, $C = -1$. Since transport currents couple only to the topological currents induced by the Chern magnetization, $I_C = M_C = C\Delta e/h$, the transport measurements at low fields display indications of $C = -1$ behavior in the vicinity of $\nu = 1$ as reported previously [19].

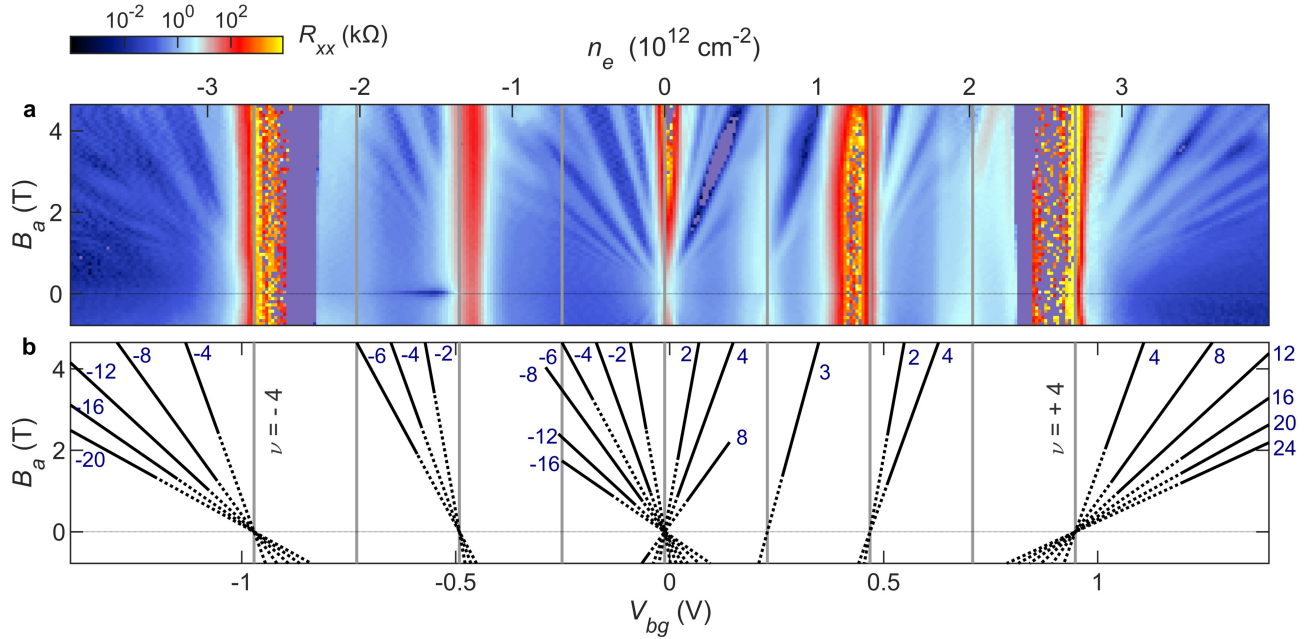

**Extended Data Fig. 2**. **Landau fan diagram. a**, Longitudinal resistance $R_{xx}$ at $T = 300$ mK reproduced from Fig. 1a. **b**, Landau level fits. Integer fillings $\nu$ of the moiré supercell are indicated by vertical grey lines and the Landau levels and Chern insulating states are indicated by diagonal lines along with their corresponding Chern numbers. Dotted lines are extrapolations.

**SOT fabrication and magnetic imaging**

The Indium SOTs were fabricated as described in Refs. [24,38] with diameters ranging from 150 to 200 nm and included an integrated shunt resistor on the tip [38]. The SOT readout was carried out using a cryogenic SQUID series array amplifier (SSAA) [50–52]. The magnetic imaging was performed in a ³He system [47] at 300 mK at which the In SOTs can operate in magnetic fields of up to 0.6 T. At fields $B_a \approx 50$ mT used in this study, the SOTs displayed flux noise down to 250 n$\Phi_0$/Hz$^{1/2}$, spin noise of 10 $\mu_B$/Hz$^{1/2}$, and field noise down to 10 nT/Hz$^{1/2}$ at frequencies above few hundred Hz. Since at lower frequencies the sensitivity is limited by $1/f$ noise, the magnetic imaging was performed in *ac* mode by applying an *ac* backgate voltage $V_{bg}^{ac}$ at a frequency of 5 to 6 kHz. For height control the SOT was attached to a quartz tuning fork as described in Ref. [53]. The tuning fork was electrically excited at the resonance frequency of ~33 kHz. The current through it was amplified using a room temperature home-built trans-impedance amplifier, designed based on Ref. [54], and measured using a lock-in amplifier. The scanning was performed at a constant height of about 160 nm above the top hBN surface. The $B_z^{ac}$ images were acquired with pixel size of 100 nm and acquisition time of 120 ms/pixel.

**Current and magnetization reconstruction**

For reconstruction of the equilibrium currents from the measured $B_z^{ac}(x, y)$ we have used the inversion method described in detail in Ref. [55]. A similar procedure was used for the magnetization reconstruction [56]. Prior to the inversion, a Gaussian high pass filter was applied to the raw data in order to remove a weak long range parasitic background arising from thermal gradients due to the backgate *ac* excitation.

Supplementary Video 2 shows the evolution of $B_z^{ac}(x, y, \nu_\uparrow)$, $B_z^{ac}(x, y, \nu_\downarrow)$, and $\Delta B_z^{ac}(x, y, \nu)$ in the range of $\nu = 0.737$ to $1.174$. The corresponding reconstructions of $m_z(x, y, \nu_\uparrow)$ and $|J^{ac}|(x, y, \nu_\uparrow)$ from $B_z^{ac}(x, y, \nu_\uparrow)$ are shown in Supplementary Video 3. A 3D tomographic representation of $m_z(x, y, \nu_\uparrow)$, $m_z(x, y, \nu_\downarrow)$, and $\Delta m_z(x, y, \nu)$ is presented in Supplementary Video 4.

**Band structure calculations**

The continuum model [1,57] is used to calculate the band structure and the orbital magnetization of MATBG. Two graphene layers are coupled by the interlayer electron hopping. For each graphene layer, lattice vectors are set as

$$\boldsymbol{a}_1 = \sqrt{3}a_0 \left(\frac{1}{2}, \frac{\sqrt{3}}{2}\right), \quad \boldsymbol{a}_2 = \sqrt{3}a_0 \left(-\frac{1}{2}, \frac{\sqrt{3}}{2}\right),$$

where $a_0 = 0.142$ nm is the intralayer bond length in graphene. The $A$ and $B$ sublattices are located at $\boldsymbol{\delta}_A = (0,0)$ and $\boldsymbol{\delta}_B = a_0(0,1)$. The corresponding reciprocal lattice vectors are

$$\boldsymbol{b}_1 = \frac{4\pi}{3a_0}\left(\frac{\sqrt{3}}{2}, \frac{1}{2}\right), \quad \boldsymbol{b}_2 = \frac{4\pi}{3a_0}\left(-\frac{\sqrt{3}}{2}, \frac{1}{2}\right),$$

and the $K$ and $K'$ valleys are located at $\boldsymbol{K}_\zeta = \zeta \frac{4\pi}{3a_0}\left(\frac{\sqrt{3}}{2}, \frac{1}{2}\right)$, where $\zeta = \pm 1$.

The Hamiltonian is given by:

$$H = H_t + H_b + H_{tb},$$

where $H_t$ and $H_b$ are the top and bottom layer Hamiltonians, given by

$$H_t = \sum_{\boldsymbol{q},\zeta,s} a_{t,s,\zeta}^\dagger(\boldsymbol{q}) \zeta \hbar v_f \widehat{\boldsymbol{R}}_+ \boldsymbol{q} \cdot \boldsymbol{\sigma} a_{t,s,\zeta}(\boldsymbol{q}),$$

$$H_b = \sum_{\boldsymbol{q},\zeta,s} a_{b,s,\zeta}^\dagger(\boldsymbol{q}) \left[\zeta \hbar v_f \widehat{\boldsymbol{R}}_- \boldsymbol{q} \cdot \boldsymbol{\sigma} + \delta \sigma_z\right] a_{b,s,\zeta}(\boldsymbol{q}).$$

Here $a_{tb,s,\zeta}^\dagger(\boldsymbol{q})$ is the electron creation operator for an electron with spin index $s = \uparrow, \downarrow$ and valley index $\zeta = \pm 1$ in the top/bottom layer, $\widehat{\boldsymbol{R}}_\pm = \cos\frac{\theta}{2} \pm i\sigma_y \sin\frac{\theta}{2}$ is the rotation matrix for top/bottom layer with twist angle $\theta$, $\boldsymbol{\sigma} = (\sigma_x, \sigma_y, \sigma_z)$ are Pauli matrices, the momentum $\boldsymbol{q}$ is defined relative to the $K_\zeta$ valley, and the Fermi velocity $\hbar v_f$ is taken to be 0.596 eV·nm. The staggered potential $\delta$, induced in the bottom graphene by the hBN substrate, breaks the inversion symmetry.

The interlayer coupling is given by

$$H_{tb} = \sum_{q,\zeta,s} a_{t,s,\zeta}^\dagger(q)\left(T_{\mathbf{q}_b,\zeta}(q,q') + T_{\mathbf{q}_{tr},\zeta}(q,q') + T_{\mathbf{q}_{tl},\zeta}(q,q')\right) a_{b,s,\zeta}(q').$$

Three interlayer hopping processes couple $q$ and $q'$, with the momentum transfer $q - q' = \{q_b, q_{tr}, q_{tl}\}$, where $q_b = \frac{8\pi\sin\frac{\theta}{2}}{3\sqrt{3}d}(-1,0)$, $q_{tr} = \frac{8\pi\sin\frac{\theta}{2}}{3\sqrt{3}d}\left(\frac{\sqrt{3}}{2},\frac{1}{2}\right)$, $q_{tl} = \frac{8\pi\sin\frac{\theta}{2}}{3\sqrt{3}d}\left(-\frac{\sqrt{3}}{2},\frac{1}{2}\right)$. The interlayer hopping matrix is given by

$$T_{\mathbf{q}_b,\zeta}(q,q') = \frac{t}{3}\begin{pmatrix} w & 1 \\ 1 & w \end{pmatrix}\delta_{q-q',q_b}$$

$$T_{\mathbf{q}_{tr},\zeta}(q,q') = \frac{t}{3}\begin{pmatrix} w & e^{-i\zeta\frac{2\pi}{3}} \\ e^{i\zeta\frac{2\pi}{3}} & w \end{pmatrix}\delta_{q-q',q_{tr}}$$

$$T_{\mathbf{q}_{tl},\zeta}(q,q') = \frac{t}{3}\begin{pmatrix} w & e^{i\zeta\frac{2\pi}{3}} \\ e^{-i\zeta\frac{2\pi}{3}} & w \end{pmatrix}\delta_{q-q',q_{tl}}$$

The interlayer hopping strength $t$ is taken to be 0.33 eV. Since MATBG is actually corrugated in the out-of-plane direction and the interlayer distance in the $AA$ stacking region is substantially increased, this lattice relaxation is modelled by the ratio $w$ of the tunneling strength between $AA$ and $AB$ regions. The reduced tunneling strength in $AA$ regions increases the band gap between the flat bands and the dispersive bands. Finally, the Hamiltonian is given by

$$H = \sum_{q,s,\zeta} A_{s,\zeta}^\dagger h_\zeta(\mathbf{q}) A_{s,\zeta}(\mathbf{q}),$$

where $A_{s,\zeta}(q)$ represent a series of states: $a_{b,s,\zeta}(q)$ and $a_{t,s,\zeta}(q')$, with $q' - q = q_b, q_{tr}, q_{tl}$. Truncating $A_{s,\zeta}(q)$ in the first shell, $A_{s\zeta}(q) = [a_{b,s,\zeta}(q), a_{t,s,\zeta}(q + \zeta q_b), a_{t,s,\zeta}(q + \zeta q_{tr}), a_{t,s,\zeta}(q + \zeta q_{tl})]$, results in

$$h_\zeta(q) = \begin{pmatrix} h_{b,\zeta}(q) & T_{\zeta,q_b} & T_{\zeta,q_{tr}} & T_{\zeta,q_{tl}} \\ T_{\zeta,q_b}^\dagger & h_{t,\zeta}(q+\zeta q_b) & 0 & 0 \\ T_{\zeta,q_{tr}}^\dagger & 0 & h_{t,\zeta}(q+\zeta q_{tr}) & 0 \\ T_{\zeta,q_{tl}}^\dagger & 0 & 0 & h_{t,\zeta}(q+\zeta q_{tl}) \end{pmatrix}.$$

The staggered potential $\delta$ is applied to the bottom graphene. As a result, the sublattice polarizations in the top and bottom layers, $P_t$ and $P_b$, are not identical. They are given by the expectation value of polarization operator $\hat{\tau}_t$ and $\hat{\tau}_b$. In the above truncated basis they are expressed as

$$\hat{\tau}_t = \begin{pmatrix} \tau_z & 0 & 0 & 0 \\ 0 & 0 & 0 & 0 \\ 0 & 0 & 0 & 0 \\ 0 & 0 & 0 & 0 \end{pmatrix}, \hat{\tau}_b = \begin{pmatrix} 0 & 0 & 0 & 0 \\ 0 & \tau_z & 0 & 0 \\ 0 & 0 & \tau_z & 0 \\ 0 & 0 & 0 & \tau_z \end{pmatrix}$$

where $\tau_z$ is the Pauli matrix in the sublattice space, $A/B$.

Figure 4a shows the resulting structure of the $KA$ flat band. The sublattice polarizations $P_t$ and $P_b$ are presented in Extended Data Fig. 3. Note that the sublattice polarization is only partial and is more pronounced in the bottom graphene layer to which the staggered potential is applied. The sublattice occupation weight $W_{A/B}$ in each layer is given by $(1 \pm P)/2$.

In this work, we have used 42 sites in the reciprocal space to construct the Hamiltonian, which gives a 84×84 Hamiltonian for a single flavor $K$. The mini Brillouin zone is sampled by a 400×400 mesh gird with additional denser grid around the $\Gamma$ point where energy bands are more dispersive [1,57].

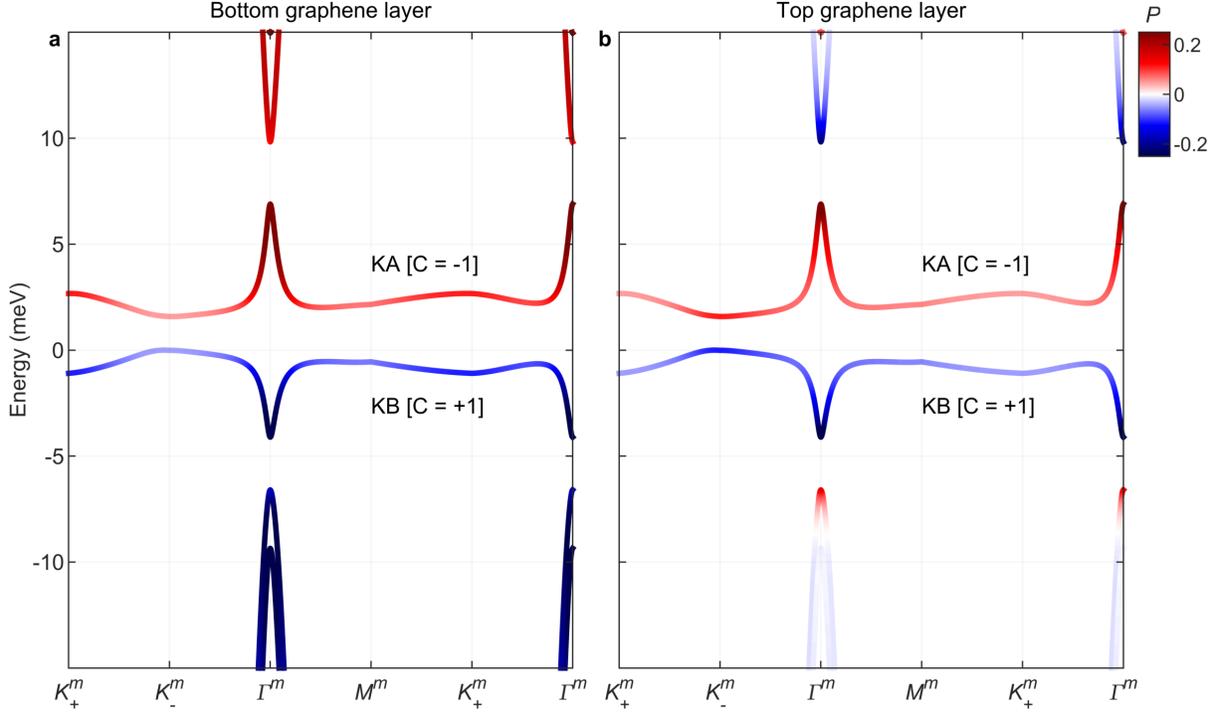

**Extended Data Fig. 3**. **Sublattice polarization. a-b**, Single particle band structure of the $K$ valley with color indicating the degree of sublattice polarization $P$ in the bottom (**a**) and top (**b**) graphene layers. Red (blue) color shows polarization of $A$ ($B$) sublattice with $P = 0.2$ corresponding to 60% occupation weight on $A$ and 40% on $B$ sublattices. The sign of $P$ determines the sign of $C$. Staggered potential $\delta = 17$ meV is applied to the bottom graphene, $\theta = 1.08°$, and $w = 0.95$.

**Orbital magnetization calculations**

The orbital magnetic dipole moment of a Bloch wave packet centered around wave vector $\boldsymbol{k}$ in band $n$ has two contributions [25,27,35–37] arising from the self-rotation of the wave packet, $m_{SR}$, and the center of mass motion of the wave packet, $m_C$, which we refer to as Chern magnetic moment because it acquires its main contribution when the chemical potential $\mu$ resides in the Chern gap:

$$\boldsymbol{m}_n^{SR}(\boldsymbol{k}) = -i\frac{e}{2\hbar}\langle \nabla_{\boldsymbol{k}} n| \times [H(\boldsymbol{k}) - \varepsilon(\boldsymbol{k})]|\nabla_{\boldsymbol{k}} n\rangle = -\frac{e}{\hbar}\text{Im}\sum_{m\neq n}\frac{\langle n|\partial_{k_x}H|m\rangle\langle m|\partial_{k_y}H|n\rangle}{\varepsilon_n(\boldsymbol{k}) - \varepsilon_m(\boldsymbol{k})}\hat{\boldsymbol{z}},$$

$$\boldsymbol{m}_n^C(\boldsymbol{k}) = -2\frac{e}{\hbar}\text{Im}\sum_{m\neq n}\frac{\langle n|\partial_{k_x}H|m\rangle\langle m|\partial_{k_y}H|n\rangle}{[\varepsilon_n(\boldsymbol{k}) - \varepsilon_m(\boldsymbol{k})]^2}[\mu - \varepsilon_n(\boldsymbol{k})]\hat{\boldsymbol{z}},$$

with the total moment

$$\boldsymbol{m}_n^{\text{tot}}(\boldsymbol{k}) = \boldsymbol{m}_n^{SR}(\boldsymbol{k}) + \boldsymbol{m}_n^C(\boldsymbol{k}) = -\frac{e}{\hbar}\text{Im}\sum_{m\neq n}\frac{\langle n|\partial_{k_x}H|m\rangle\langle m|\partial_{k_y}H|n\rangle}{[\varepsilon_n(\boldsymbol{k}) - \varepsilon_m(\boldsymbol{k})]^2}\left[(\varepsilon_n(\boldsymbol{k}) - \varepsilon_m(\boldsymbol{k})) + 2(\mu - \varepsilon_n(\boldsymbol{k}))\right]\hat{\boldsymbol{z}}.$$

Here $|n\rangle$ (shorthand for $|n(\mathbf{k})\rangle$) is the spatially-periodic part of the Bloch eigenstate of band $n$ with energy $\varepsilon_n(\mathbf{k})$ and momentum $\mathbf{k}$, and $\partial_k H$ is the velocity operator. The expression clearly shows that both contributions to $\mathbf{m}_n(\mathbf{k})$ originate from the Berry curvature $\mathbf{\Omega}_n(\mathbf{k})$,

$$\mathbf{\Omega}_n(\mathbf{k}) = -2\,\mathrm{Im} \sum_{m \neq n} \frac{\langle n|\partial_{k_x} H|m\rangle \langle m|\partial_{k_y} H|n\rangle}{[\varepsilon_n(\mathbf{k}) - \varepsilon_m(\mathbf{k})]^2} \hat{\mathbf{z}}.$$

The total orbital magnetization of band $n$ is given by

$$\mathbf{M}_n^{\mathrm{tot}} = -\frac{e}{\hbar} \int \frac{d^2k}{(2\pi)^2} f(\varepsilon_n(\mathbf{k})) \mathrm{Im} \sum_{m \neq n} \frac{\langle n|\partial_{k_x} H|m\rangle \langle m|\partial_{k_y} H|n\rangle}{[\varepsilon_n(\mathbf{k}) - \varepsilon_m(\mathbf{k})]^2} [(\varepsilon_n(\mathbf{k}) - \varepsilon_m(\mathbf{k})) + 2(\mu - \varepsilon_n(\mathbf{k}))] \hat{\mathbf{z}},$$

where $f(\varepsilon_n(\mathbf{k}))$ is the Fermi-Dirac distribution function and $\mu$ is the chemical potential. Because of the denominator, near the top of the band $n$, most of the contribution to the magnetization originates from virtual transitions to the closest-in-energy band above it, $m = n + 1$. Considering only these two bands, if the chemical potential $\mu$ resides in the gap between the bands, $\varepsilon_n(\mathbf{k}) < \mu < \varepsilon_{n+1}(\mathbf{k})$, it is easy to see that $M_{SR}$ and $M_C$ should typically be of opposite sign. Moreover, in the presence of particle-hole symmetry, $\varepsilon_n(\mathbf{k}) = -\varepsilon_{n+1}(\mathbf{k})$, the expression for $M$ reduces to [36]

$$\mathbf{M} = \frac{2e}{\hbar} \int \frac{d^2k}{(2\pi)^2} f(\varepsilon(\mathbf{k})) \Omega(\mathbf{k}) \mu\, \hat{\mathbf{z}},$$

highlighting further the Berry curvature origin of both contributions to the orbital magnetization.

The $M_{SR}$ of a valley-polarized flat band in MATBG has been recently derived numerically using single particle band structure calculations [27]. Here, we carry out a similar calculation of $M_{SR}$ and additionally derive the $M_C$. The Chern number at a given $\mu$ is determined by the sum of the Chern numbers of all the occupied bands. In our single particle calculations, we assume from symmetry considerations that $M_{SR}$, $M_C$, $m_{SR}$, $m_C$, and $C$ are all equal to zero at CNP. We then calculate numerically the $m_{SR}(\mathbf{k})$ of a single $KA$ polarized conduction band for $\theta = 1.08°$, $\delta = 17$ meV, and $w = 0.95$ as shown in Fig. 4b. The mini Brillouin zone is sampled by a 400×400 mesh gird with additional denser grid around the $\Gamma$ point. A similar calculation is performed for $m_C(\mathbf{k})$.

The $m_{SR}(\mathbf{k})$ and $m_C(\mathbf{k})$ are then integrated up to $\mu$, and the resulting $M_{SR}$ and $M_C$ are plotted in Fig. 4c vs. the filling factor $\nu$ in the compressible state and vs. $\mu$ in the gap. The interaction induced Chern gap size of $\Delta = 7$ meV, comparable to the experimentally evaluated value, was introduced manually with a constant $M_{SR}$ and linear $M_C$ with a slope of $dM_C/d\mu = Ce/h$ with $C = -1$. The corresponding differential magnetization, equivalent to the signal measured in the experiment, is shown in Fig. 4d with $m_{SR} = dM_{SR}/dn$ and $m_C = dM_C/dn$ in the compressible regions and $m_{SR} = dM_{SR}/d\mu$ and $m_C = dM_C/d\mu$ in the gap. The following band at $\nu > 1$ is assumed to be of the opposite valley and thus have the same $m_{SR}$ and $m_C$ evolution, but of opposite sign.

We have explored numerically the dependence of the magnetization on the various parameters in the range $1.0° \leq \theta \leq 1.2°$, $9 \leq \delta \leq 29$ meV, and $0.75 \leq w \leq 1.00$. Although the qualitative behavior remains similar in this range of parameters, there are significant quantitative differences, which are particularly pronounced for variations in the tunneling strength ratio $w$. Extended Data Fig. 4 shows the results attained for the same $\theta = 1.08°$ and $\delta = 17$ meV as in Fig. 4, but with a lower $w = 0.8$ [58]. The $M_{SR}$ and $M_C$ values are very similar, albeit slightly lower than in Fig. 4c, and remain consistent with the experimental values in Figs. 5b,c. The main difference is observed in $m_{SR}$ and $m_C$, which show a more gradual drop towards negative values close to the top of the band. As a result, $m_z = m_{SR} + m_C$ shows a weaker maximum value $m_z^{peak}$ near $\nu \cong 0.9$. Extended Data Fig. 4d summarizes the values of $m_z^{peak}$ calculated for a range of $\delta$ and $w$. The simulation results for $w \gtrsim 0.85$ are consistent with experimental values of $m_z^{peak} \gtrsim 5\ \mu_B/e$. Note that the calculations are based on the assumption of flavor degeneracy lifting from CNP. If the degeneracy lifting occurs only above a certain filling factor $\nu_{dl}$ [5–7], $M_z$ will remain zero below $\nu_{dl}$ and will raise faster above $\nu_{dl}$, giving rise to larger $m_z$ and $m_z^{peak}$ on approaching the top of the band.

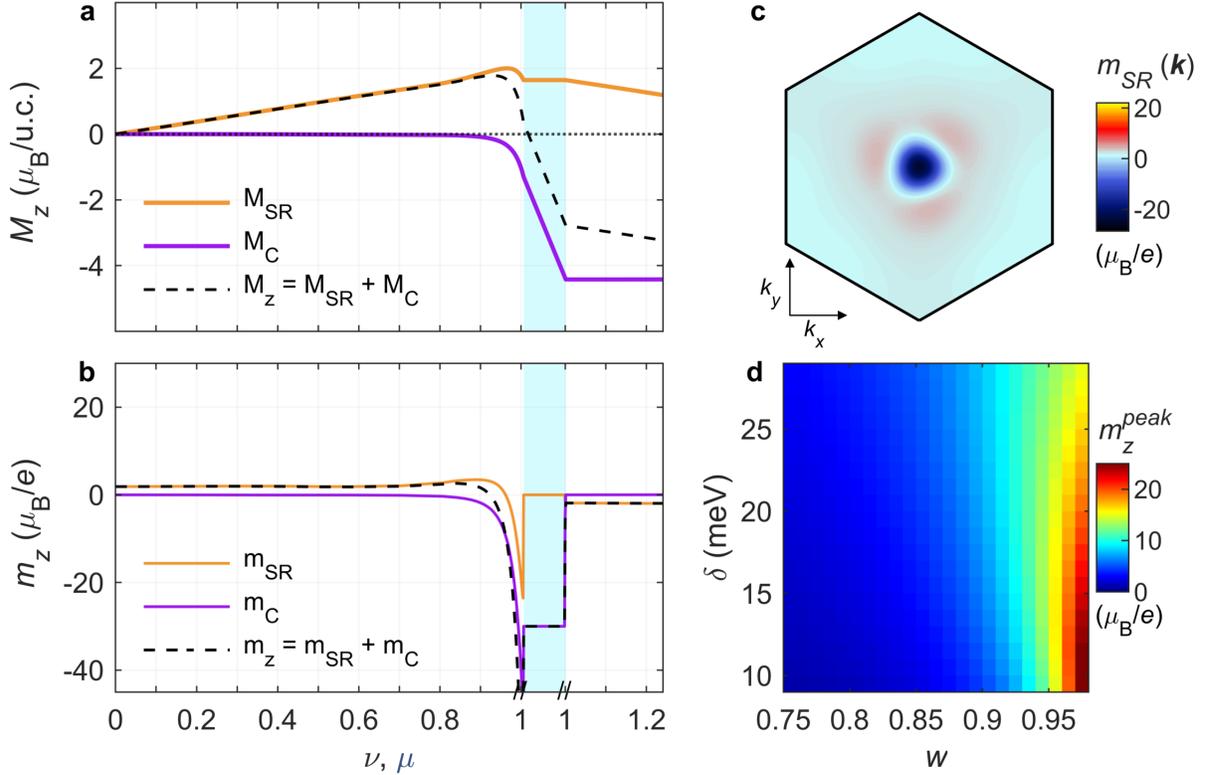

**Extended Data Fig. 4**. **Orbital magnetization calculations. a,** The evolution of the integrated magnetization $M_z$ (dashed) and its two components $M_{SR}$ and $M_C$ vs. the filling factor $\nu$ in the compressible region and vs. $\mu$ in the gap region (cyan) in the $KA$ flat band for twist angle $\theta = 1.08°$, staggered potential $\delta = 17$ meV, and tunneling ratio $w = 0.8$. **b,** The corresponding evolution of the differential magnetization $m_z$ and its $m_{SR}$ and $m_C$ components. **c,** The self-rotation magnetization $m_{SR}(\mathbf{k})$ in the first mini-Brillouin zone in the $KA$ flat band. **d,** The calculated peak value of the differential magnetization $m_z^{peak}$ near the top of the band as a function of $\delta$ and $w$ for $\theta = 1.08°$.

An important insight from the above calculations is that the Berry-curvature-induced magnetization is a sensitive probe of the local band structure. We note, however, that the calculation of the magnetization was done for non-interacting electrons, and does not include the effect of interactions. Moreover, the calculations do not take into account strain, which can be expected to modify the magnetism further [27]. The large spatial variations in the local magnetization revealed in Supplementary Video 4 point out the complexity of the local band structure and of the substrate potential.

## Evolution of $m_z$ across the $KA$ to $K'A$ recondensation phase transition

We inspect the behavior of $m_z = m_{SR} + m_C$ more closely by analyzing its evolution along the dashed yellow line in Fig. 4g as shown in Extended Data Fig. 5b. The $m_z$ shows two sign changes as a function of $\nu$. The first occurs at $\nu_0$ (dotted black line in Fig. 4g) that marks the transition from $m_{SR}$ to $m_C$ dominated magnetization. This transition is continuous and thus $m_z$ goes smoothly through zero and is additionally smeared by our finite $ac$ modulation of $\nu^{ac} = 0.083$. This sign change occurs slightly below $\nu = 1$, as shown by the numerical calculations in Fig. 4d and Extended Data Fig. 5a, and its position depends on the band structure parameters.

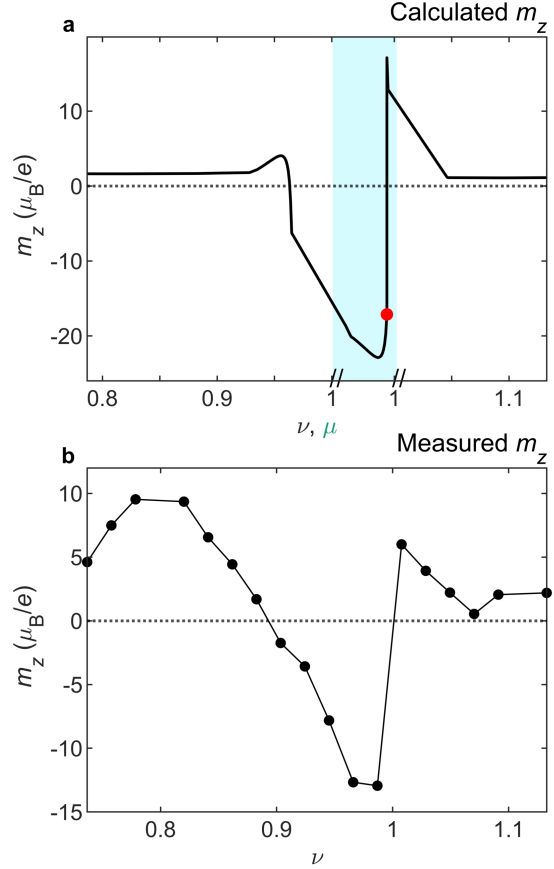

**Extended Data Fig. 5**. **Evolution of $m_z$ through recondensation phase transition. a,** Numerically calculated $m_z$ in the vicinity of $\nu = 1$, similar to Fig. 4d, but including the first-order recondensation transition marked by the red dot, upon which the $m_z$ flips its sign discontinuously. The value of the Chern gap is taken to be 3.5 meV and $m_z = dM_z/dn$ is calculated taking into account the finite $\nu^{ac} = 0.083$ modulation. **b,** The measured $m_z$ along the vertical yellow dashed line in Fig. 4g. The black dots are the experimental data points and the solid curve is guide to the eye. The first sign change of $m_z$ at $\nu \cong 0.9$ marks the continuous crossover from $m_{SR}$ to $m_C$ dominated magnetization. The second sign change at $\nu \cong 1$ is discontinuous (and hysteretic) and reflects the first-order $K$ to $K'$ recondensation transition.

The second sign change occurs at $\nu_{\uparrow f}$ (solid black line in Fig. 4g) due to completely different mechanism of carrier recondensation from valley $K$ to $K'$ through a first-order transition. As shown in Fig. 4c, the total magnetization $M_z$ decreases with increasing the chemical potential in the gap region due to the negative contribution of $M_C$ ($dM_C/d\mu = Ce/h$ with $C = -1$). If the gap is sufficiently large, the $M_z$ will become negative (red dot in Fig. 4c). As a result, it becomes energetically favorable to recondense all the carriers from valley $K$ to $K'$, resulting in a discontinuous flipping of the magnetization. Figures 4c,d show the expected evolution of the magnetization in absence of this flipping transition, while Extended Data Fig. 5a presents the calculated $m_z$ with the flipping. Since this transition is discontinuous and hysteretic, it is not smeared by the finite $v^{ac}$, and thus remains discontinuous also in the experimental data as indicated by the black line in Fig. 4g and by the jump between the neighboring experimental data points in Extended Data Fig. 5b.

Note that in the absence of the recondensation transition, for a $C = -1$ patch, the carriers first fill the $KA$ band followed by filling of the $K'A$ band (ignoring spin), as indicated in Fig. 4c. In presence of the transition, however, the sequence is different. The $KA$ band is filled out first, then as the chemical potential moves through the gap, the $KA$ band is emptied abruptly and the $K'A$ is filled completely through a first-order transition, followed by gradual filling of the $KA$ band again.

### hBN alignment and moiré commensurability

As discussed above, in our sample the graphene is apparently not aligned with hBN. However, according to theoretic analysis in [20], formation of a Chern mosaic on µm scale does not require alignment of the atomic lattices. Instead, the relevant consideration is the commensurability of the two moiré lattices, one arising from the twist angle $\theta_{GG}$ between the two graphene sheets, and the other determined by the twist angle $\theta_{GBN}$ between hBN and the adjacent graphene. The conditions for such commensurability of the two moiré lattices are given by the following equations for $\theta_{GG}, \theta_{GBN}$ [20]:

$$\theta_{GG}^{\pm} = \arccos\frac{t}{\sqrt{t^2 + s^2}} \pm \arccos\frac{t + \frac{1}{2}(1 - \frac{1}{\alpha^2})}{\sqrt{t^2 + s^2}},$$

$$\theta_{GBN}^{\pm} = \arccos\frac{r - 1}{\sqrt{(r-1)^2 + s^2}} \pm \arccos\frac{\alpha r - \frac{1}{2}(\alpha + \frac{1}{\alpha})}{\sqrt{(r-1)^2 + s^2}},$$

where $\alpha = 1.017$ is the ratio between the hBN and graphene lattice constants, $t = r^2 + s^2 - r$, $s = \sqrt{3}(\frac{p}{n} + \frac{q}{n})/2$, $r = (\frac{p}{n} - \frac{q}{n})/2$, and $n, p$, and $q$ are triplets of coprime integers. The resulting super moiré cell is $n^2$ times larger than the $\theta_{GG}$ moiré cell, thus we only consider low values of $n$ ($n = 1,2$). Solutions to the above equations are shown in Extended Data Fig. 6 for values of $\theta_{GG}$ close to the magic angle. Numerous additional solutions exist for larger $n$. Note that due to the symmetry of the system, each $\theta_{GG}, \theta_{GBN}$ solution has six-fold rotational symmetry. Furthermore, it was shown that the $\theta_{GG}, \theta_{GBN}$ pair can vary some amount from the commensurate angle (depending on the value of $n$), implying that there is a finite width around each solution that will lead to a Chern mosaic, as indicated by the error bars in Extended Data Fig. 6. Additionally, it was pointed out in [20] that the commensurate angles probably create a lower energy state, implying that a sample might relax into these states. Finally, taking into account the possible commensurability with either top or bottom hBN, presence of strain, charge disorder, $\theta_{GG}$ disorder, and presumably $\theta_{GBN}$ disorder, the formation of moiré commensurability might be ubiquitous. In particular, the $\theta_{GBN}$ values of 8° and 6° of the top and bottom hBN evaluated from the optical images in Extended Data Fig. 1 are within the range of possible angles for formation of commensurate moiré lattices in Extended Data Fig. 6.

Note that the above analysis ignores all types of disorder and specifically twist angle disorder which is known to be significant in MATBG samples [40]. A more realistic picture includes twist angle disorder in both $\theta_{GG}, \theta_{GBN}$ which requires a coarse-grained analysis leading to patches that are stabilized by disorder combined with average substrate potential.

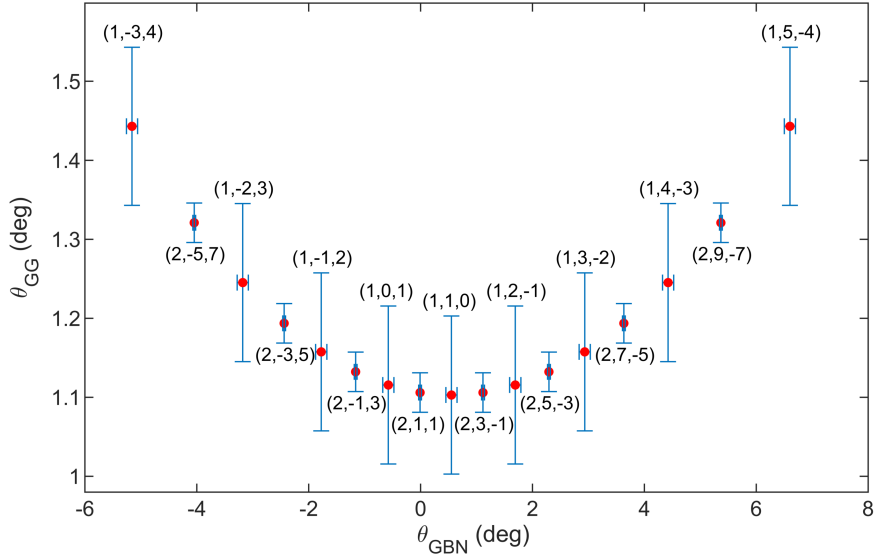

**Extended Data Fig. 6**. **Moiré commensurability conditions**. Calculated pairs of graphene-graphene $\theta_{GG}$ and graphene-hBN $\theta_{GBN}$ twist angles that result in commensurability of the two moiré lattices following Ref. [20]. Each solution is labeled by its ($n, p, q$) integer triplet. Shown are only solutions for $n = 1,2$ as the super moiré cell scales like $n^2$, but numerous additional solutions exist for larger values of $n$ with proportionally larger super moiré cells. The blue error bars show deviations from the exact commensurate angles where a Chern mosaic structure is still expected to form.

**Experimental derivation of the Chern mosaic**

The differential magnetization $m_z(x, y, \nu)$ is a continuous function of $\nu$, except at the abrupt flipping of the magnetization due to the recondensation of carriers from $K$ to $K'$ valley. We are interested in determining the filling factor $\nu_o(x, y)$, where $m_z(x, y)$ traverses zero continuously upon crossing from $m_{SR}(x, y)$ to $m_C(x, y)$ dominated magnetization as $\mu$ approaches the Chern gap, and the filling factors $\nu_{\uparrow f}(x, y)$ and $\nu_{\downarrow f}(x, y)$ where $m_z(x, y)$ flips its sign discontinuously. The finite width of our *ac* modulation of the filling factor ($\nu^{ac} = 0.083$ peak to peak), causes some averaging of $m_z(x, y, \nu)$ over the $\nu^{ac}$ range. This narrow averaging has no essential effect on $\nu_o(x, y)$, since $m_z$ is continuous there. It also has no appreciable effect on determining $\nu_{\uparrow f}$ or $\nu_{\downarrow f}$ in presence of hysteresis: if the hysteresis width is larger than $\nu^{ac}$ the observed transition will remain discontinuous because once the carriers have recondensed for from $K$ to $K'$ they cannot flip their valley polarization back within the *ac* cycle of $\nu^{ac}$ modulation.

To determine $\nu_{\uparrow f}(x, y)$ we calculate $\delta m_z(x, y, \nu_{\uparrow})$ by subtracting consecutive images. Due to the discontinuous jump, $|\delta m_z(x, y, \nu_{\uparrow})|$ typically attains its maximal value at $\nu_{\uparrow f}(x, y)$, which we then confirm by inspection. Similar procedure is used for mapping $\nu_{\downarrow f}(x, y)$. Determination of $\nu_o(x, y)$ is attained by tracking the minimum $|m_z(x, y)|$ below $\nu_{\uparrow f}(x, y)$. Integrating $m_z(x, y)$ from our lowest filling factor up to $\nu_o(x, y)$, we obtain the map of $M_{SR}^{max}$ presented in Fig. 5b. Similarly, by integrating $m_z(x, y)$ from $\nu_o(x, y)$ to $\nu_{\uparrow f}(x, y)$, we obtain a map of the Chern magnetization (Fig. 5c). This is a lower bound on $M_C^{max}$, or equivalently on the Chern gap size $\Delta = M_C^{max}(h/Ce)$, since the flipping happens at some point within the gap, but not necessarily at its top. The Chern mosaic is then extracted from $m_z(x, y, \nu_{\uparrow})$ as follows: red (blue) regions in Fig. 5d correspond to positive (negative) $m_z(x, y, \nu_{\uparrow})$ at filling factors below $\nu_o(x, y)$, and green regions have $|m_z(x, y, \nu)|$ that never exceeds 5 $\mu_B/e$, indicating weak orbital magnetization over the entire range of filling factors. Note that $m_z(x,y)$ is a continuous function while $C(x,y) \in (-1,0,1)$ is discrete, resulting in sharp edges.

The Chern mosaic can be also extracted from analyzing the sign of $m_z(x,y,\nu_\uparrow)$ at filling factors between $\nu_o(x,y)$ and $\nu_{\uparrow f}(x,y)$, which reflects $m_C$ in the Chern gap, which is of opposite sign to that of $m_{SR}$. Alternatively, one can analyze the sign of the local integrated $M_{SR}$ and $M_C$, rather than the differential magnetization. All four procedures give rise to essentially identical maps of the Chern mosaic.

**Local twist angle**

In Ref. [40] a map of the twist angle $\theta(x,y)$ was derived by measurement of the local $n_s(x,y)$ by tracing the Landau levels in the dispersive bands at $B_a \cong 1$ T using a Pb SOT. In the current study an In SOT was used instead, which has a higher sensitivity at low fields but precludes imaging of Landau levels at high fields that are necessary for the detailed mapping of $\theta(x,y)$. Nevertheless, by measuring $B_z^{ac}(x)$ at $B_a = 173$ mT along the dotted line in Fig. 2b, we succeeded to trace the $n_s^+(x)$ and $n_s^-(x)$ of the electron and hole dispersive bands and the corresponding filling factors $\nu_s^\pm(x) = n_s^\pm(x)/n_s^{global}$ as shown by the black curves in Extended Data Fig. 7a (here, $n_s^{global}$ is the $n_s$ extracted from the transport data). The derived local twist angle along this line, $\theta(x) = \sqrt{\frac{\sqrt{3}}{8} a^2 \frac{n_s^+ - n_s^-}{2}}$, is shown in Extended Data Fig. 7b. The left side of the sample shows a higher $\theta$ which may be related to the observation of larger $M_C$ and the larger local hysteresis in this area of the sample in Fig. 5.

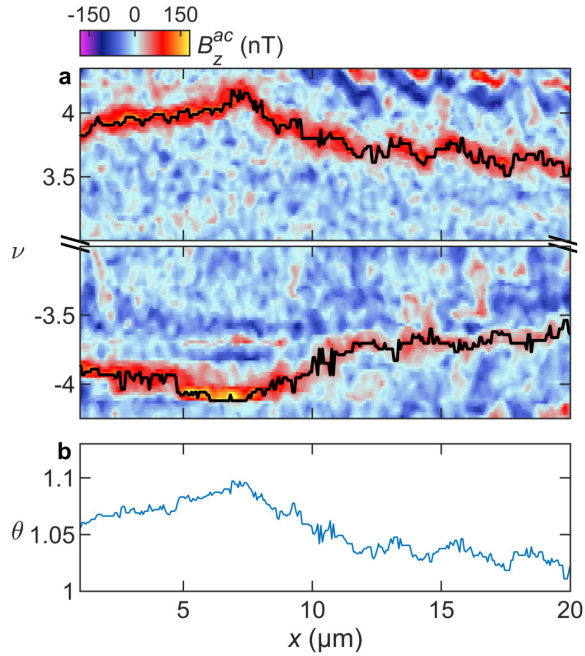

**Extended Data Fig. 7**. **Local twist angle. a,** $B_z^{ac}(x)$ measured along the dotted line in Fig. 2b vs. the global filling factor $\nu$ in the vicinity of $\nu = 4$ (top) and $\nu = -4$ (bottom). The black lines mark $\nu_s^+(x)$ and $\nu_s^-(x)$ corresponding the local electron and hole dispersive band edges, $n_s^+(x)$ and $n_s^-(x)$. **b,** The derived local twist angle $\theta(x) = \sqrt{\frac{\sqrt{3}}{8} a^2 \frac{n_s^+ - n_s^-}{2}}$.

**Comparison of magnetization and transport hysteresis**

By analyzing the $\Delta m_z(x, y, \nu)$ images, we extract the number of pixels that are locally hysteretic (above a noise threshold of 2 $\mu_B/e$). Extended Data Fig. 8a shows a histogram of the percentage of the sample area that has locally hysteretic $m_z$ at each filling factor $\nu$. For comparison, Extended Data Fig. 8b shows the transport hysteresis, $\Delta R_{yx}(\nu) = R_{yx}(\nu_\uparrow) - R_{yx}(\nu_\uparrow)$, from Fig. 1d. The magnetization and transport hysteresis have very similar shapes, starting with a slow onset, followed by a clear peak and drop off. Note that $R_{yx}$ is measured over the contacts shown in Fig. 2b, whereas the magnetization hysteresis is calculated over the entire sample. Also note that the comparison is not straightforward as $R_{yx}$ depends not only on magnetization but also on carrier density.

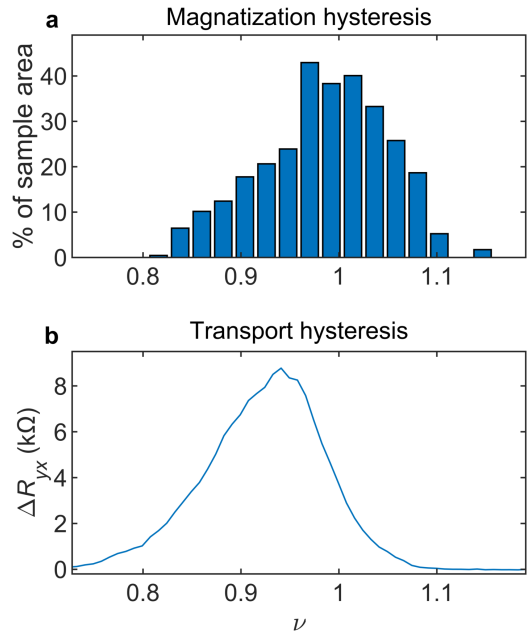

**Extended Data Fig. 8**. **Comparison of magnetization and transport hysteresis. a,** Percentage of sample area that has locally hysteretic $m_z$ represented as a histogram as a function of filling factor $\nu$. **b**, The hysteresis in transport, $\Delta R_{yx}(\nu) = R_{yx}(\nu_\uparrow) - R_{yx}(\nu_\uparrow)$, at $B_a = 47$ mT from data presented in Fig. 1d.

**Captions of Supplementary Videos**

**Supplementary Video 1 | Minor hysteresis loops of $R_{yx}$ around $\nu = 1$ at $B_a = 46$ mT**. Top panel: the system is initialized at filling factor $\nu_{low} = 0.55$ and $R_{yx}$ is measured while sweeping $\nu$ up to $\nu_{max}$ (magenta curves) and decreasing back to $\nu_{low}$ (cyan curves). The procedure is then repeated upon gradually incrementing $\nu_{max}$ up to $\nu_{high} = 1.3$. Bottom panel: minor loops performed in the opposite direction. The system is initialized at $\nu_{high} = 1.3$ and $\nu$ is decreased down to $\nu_{min}$ (cyan curves) and then increased back to $\nu_{high}$ (magenta curves), and the procedure is repeated upon decrementing $\nu_{min}$ down to $\nu_{low}$. When the system is initialized at $\nu_{low}$, reversible $R_{yx}$ is observed upon sweeping the filling factor up to $\nu_{max} \cong 1$, above which a progressively increasing hysteresis develops with sharp jumps. Similar behavior is observed upon initializing the system at $\nu_{high}$, with reversible $R_{yx}$ down to $\nu_{min} \cong 0.95$ and appearance of increasing hysteresis upon lowering $\nu_{min}$. The global $R_{yx}$ dynamics is consistent with the observation of the flipping of the magnetization domains in the local measurements and with their hysteresis in Figs. 5e,f.

**Supplementary Video 2 | Evolution of $B_z^{ac}$ with filling factor**. Top panel: the measured local *ac* magnetic field $B_z^{ac}(x, y, \nu_\uparrow)$ induced by a small *ac* modulation of the filling factor $\nu^{ac} = 0.083$ upon sweeping the *dc* filling factor from $\nu = 0.737$ to $1.174$. The system is initialized at $\nu = 0$ prior to the measurement. Middle panel: same as top panel after initializing the system at $\nu = 2$ and measuring $B_z^{ac}(x, y, \nu_\downarrow)$ upon sweeping $\nu$ down from $1.174$ to $0.737$. Bottom panel: numerical difference, $\Delta B_z^{ac}(x, y, \nu)$, of the sweep up and sweep down data revealing the local hysteresis in $B_z^{ac}$. The movie shows a complex pattern of positive and negative $B_z^{ac}$, where different areas of the sample develop hysteresis at different filling factors, with some areas showing completely reversible behavior throughout the full range of $\nu$. The left side of the sample shows local hysteresis over a larger range of $\nu$.

**Supplementary Video 3 | Evolution of magnetization and equilibrium currents with filling factor**. Numerical reconstruction of the magnitude of the equilibrium current $\boldsymbol{J}^{ac}(x, y, \nu_\uparrow)$ (top panel) and of the local magnetization $m_z(x, y, \nu_\uparrow) = dM_z(x, y, \nu_\uparrow)/dn$ (bottom panel) obtained by the inversion of $B_z^{ac}(x, y, \nu_\uparrow)$ for $\nu = 0.737$ to $1.174$. In 2D the magnetization $M$ (magnetic dipole moment per unit area) is given in units of current (A), which for the case of a uniformly magnetized domain describes the equilibrium current that circulates along the edges of the domain. Thus $m_z(x, y)$ and $\boldsymbol{J}^{ac}(x, y)$ present equivalent descriptions of the change in magnetization and in equilibrium currents, respectively, due to a change in filling factor. At low $\nu$ isolated patches of positive $m_z(x, y, \nu_\uparrow)$ are present along with areas of negative response, but at higher $\nu$ the system develops into an intricate pattern of positive and negative patches and an equivalently intricate current network.

**Supplementary Video 4 | Tomographic rendering of the magnetization**. Evolution of $m_z(x, y, \nu_\uparrow)$, $m_z(x, y, \nu_\downarrow)$, and $\Delta m_z(x, y, \nu)$ as a function of the spatial axes $x$ and $y$ and the filling factor $\nu$. Different two-dimensional slices in the $x - \nu$, $y - \nu$, and $x - y$ planes are presented sequentially. The bottom panel reveals a wider hysteresis in $\nu$ in the left part of the sample and very narrow or no hysteresis in the right part.